\documentclass[%
 reprint,
superscriptaddress,
nofootinbib,
 amsmath,amssymb,
 onecolumn,
prd,
]{revtex4-2}

\usepackage{graphicx}
\usepackage{dcolumn}
\usepackage{bm}
\usepackage[colorlinks]{hyperref}

\usepackage[table,svgnames]{xcolor}
\usepackage{booktabs}

\definecolor{vlightgray}{gray}{0.9}
\usepackage{diagbox}
\usepackage{color}
\usepackage[caption=false]{subfig}

\newcommand{\bv}[1]{\boldsymbol{#1}}

\newcommand{\hMpc}{h\,\mathrm{Mpc}^{-1}}

\newcommand{\av}[1]{\left\langle{#1}\right\rangle} 

\newcommand{\vk}{\bv k}

\newcommand{\vp}{\vec p}

\newcommand{\vd}{\vec d}

\renewcommand{\vr}{\bv r}

\newcommand{\C}{\mathsf{C}}

\newcommand{\B}{\mathsf{B}}

\newcommand{\N}{\mathsf{N}}
\newcommand{\Ci}{\mathsf{C}^{-1}}

\newcommand{\Tr}[1]{\operatorname{Tr}\left[{#1}\right]}

\newcommand{\be}{\begin{equation}}
\newcommand{\ee}{\end{equation}}

\newcommand{\hun}{\,\mathrm{km}\,\mathrm{s}^{-1}\mathrm{Mpc}^{-1}}

\newcommand{\astfootnote}[1]{%
\let\oldthefootnote=\thefootnote%
\setcounter{footnote}{0}%
\renewcommand{\thefootnote}{\fnsymbol{footnote}}%
\footnote{#1}%
\let\thefootnote=\oldthefootnote%
}

\usepackage{color}
\def\beq{\begin{eqnarray}}
\def\eeq{\end{eqnarray}}

\DeclareSymbolFont{toneletters}{T1}{\familydefault}{m}{it}
\SetSymbolFont{toneletters}{bold}{T1}{\familydefault}{bx}{it}

\DeclareMathSymbol\edth{\mathord}{toneletters}{"F0}

\usepackage{empheq}

\newcommand{\resub}[1]{#1}

\let\vec\mathbf

\begin{document}


\title{\Large The BOSS DR12 Full-Shape Cosmology:\\
\normalsize $\Lambda$CDM Constraints from the Large-Scale Galaxy Power Spectrum and Bispectrum Monopole}

\author{Oliver H.\,E. Philcox}
\email{ohep2@cantab.ac.uk}
\affiliation{Department of Astrophysical Sciences, Princeton University,\\ Princeton, NJ 08540, USA}%
\affiliation{School of Natural Sciences, Institute for Advanced Study, 1 Einstein Drive,\\ Princeton, NJ 08540, USA}
\author{Mikhail M. Ivanov}\email[Einstein Fellow,\ ]{ivanov@ias.edu}
\affiliation{School of Natural Sciences, Institute for Advanced Study, 1 Einstein Drive,\\ Princeton, NJ 08540, USA}


\begin{abstract}
We present a full $\Lambda$CDM analysis of the BOSS DR12 dataset, including information from the power spectrum multipoles, the real-space power spectrum, the reconstructed power spectrum and the bispectrum monopole. This is the first analysis to feature a complete treatment of the galaxy bispectrum, including a consistent theoretical model and without large-scale cuts. Unlike previous works, the statistics are measured using window-free estimators: this greatly reduces computational costs by removing the need to window-convolve the theory model. Our pipeline is tested using a suite of high-resolution mocks and shown to be robust and precise, with systematic errors far below the statistical thresholds. Inclusion of the bispectrum yields consistent parameter constraints and shrinks the $\sigma_8$ posterior by $13\%$ to reach $<5\%$ precision; less conservative analysis choices would reduce the error-bars further. Our constraints are broadly consistent with \textit{Planck}: in particular, we find $H_0 = 69.6^{+1.1}_{-1.3}\,\mathrm{km}\,\mathrm{s}^{-1}\mathrm{Mpc}^{-1}$,  $\sigma_8 = 0.692^{+0.035}_{-0.041}$ and $n_s=0.870^{+0.067}_{-0.064}$, including a BBN prior on the baryon density. When $n_s$ is set by \textit{Planck}, we find $H_0 = 68.31^{+0.83}_{-0.86}\,\mathrm{km}\,\mathrm{s}^{-1}\mathrm{Mpc}^{-1}$ and $\sigma_8 = 0.722^{+0.032}_{-0.036}$. Our $S_8$ posterior, $0.751\pm0.039$, is consistent with weak lensing studies, but lower than \textit{Planck}. Constraints on the higher-order bias parameters are significantly strengthened from the inclusion of the bispectrum, and we find no evidence for deviation from the dark matter halo bias relations. These results represent the most complete full-shape analysis of BOSS DR12 to-date, and the corresponding spectra will enable a variety of beyond-$\Lambda$CDM analyses, probing phenomena such as the neutrino mass and primordial non-Gaussianity.
\end{abstract}

\maketitle

\section{Introduction}\label{sec: intro}


In the standard paradigm, the distribution of matter in the early Universe obeys Gaussian statistics, and can thus be fully described by its power spectrum \citep[e.g.,][]{Starobinsky:1982ee,1982PhLB..116..335L}. As the Universe evolves, gravitational evolution induces non-Gaussianity \citep[e.g.,][]{2002PhR...367....1B}, affording a complex matter distribution, with information distributed over a broad range of statistics. Why, then, do so few analyses of late-time cosmological data \citep[e.g.,][]{Scoccimarro:2000sp,Sefusatti:2006pa,2015MNRAS.451..539G,2017MNRAS.465.1757G}, use observables beyond the power spectrum?

Since the early 2000s, spectroscopic galaxy surveys have been played a role in the development of the cosmological model through their ability to measure the Universe's growth rate and expansion history using the baryon acoustic oscillation (BAO) feature \citep[e.g.,][]{2015MNRAS.451..539G,2021JCAP...11..031B,2017MNRAS.465.1757G,2020MNRAS.498.2492G,2017MNRAS.470.2617A,2017MNRAS.464.3409B,2017MNRAS.466.2242B}. This provides a `standard ruler' whose physical scale can be predicted by theory, and whose angular scale can be measured observationally, giving information on the distance-redshift relation, and hence the Hubble parameter $H(z)$ and angular diameter distance $D_A(z)$. Due to gravitational evolution, BAO effects are present not only in the power spectrum but also higher-order statistics: usually, one accounts for this by performing BAO reconstruction \citep[e.g.,][]{2007ApJ...664..675E,2009PhRvD..80l3501N,2012MNRAS.427.2132P} to shift the corresponding information back into the two-point function \citep{2015PhRvD..92l3522S} (though see \citep{2018MNRAS.478.4500P} for a BAO measurement using the bispectrum).

Whilst the BAO is certainly an important part of the galaxy power spectrum, it is by no means the only feature. Recent developments in theoretical modeling, in particular, the development of the Effective Field Theory of Large Scale Structure (EFTofLSS), have allowed a number of groups to fit the full power spectrum shape (rather than just the oscillatory component), and thus place constraints on the cosmological parameters directly \citep{2020JCAP...05..042I,2020JCAP...05..005D,2021arXiv211005530C,2021arXiv211006969K}. This is akin to the analysis of CMB data, and enhances the cosmological utility of current and future surveys. The approach has further been combined with BAO information from reconstructed spectra \citep{2020JCAP...05..032P}, and used to place constraints on non-standard cosmological models, such as non-flat universes \citep{2021PhRvD.103b3507C}, Early Dark Energy \citep{2020PhRvD.102j3502I,2021JCAP...05..072D}, massive relics \citep{2021arXiv210709664X} and ultralight axions \citep{Lague:2021frh}. Furthermore, it can be used to extract quantities orthogonal to that found in traditional scaling analyses, such as the scale of matter-radiation equality \citep{2021PhRvD.103b3538P}, which provides a powerful test of the cosmological model.

The natural question is whether such analyses can be extended to higher-order statistics, such as the galaxy bispectrum. Indeed, this is a natural place to look, since the bispectrum appears at the same order in perturbation theory as the oft-used one-loop power spectrum. However, modeling the bispectrum is difficult. Although a wealth of previous studies have discussed its theoretical form, both for matter and galaxies \citep{1998ApJ...496..586S,1999ApJ...517..531S,2000ApJ...544..597S,2001MNRAS.325.1312S,2015JCAP...10..039A,2015JCAP...05..007B,2020arXiv201100899K,2021JCAP...07..008G,2021JCAP...04..029H,2021JCAP...03..021A,2021JCAP...03..105B,2015JCAP...10..039A,Eggemeier:2018qae,Byun:2020rgl,Eggemeier:2021cam,Oddo:2021iwq,Osato:2021nza,Taruya:2021ftd,Baldauf:2021zlt}, it is only recently that we have obtained a theory model that is capable of predicting the full galaxy bispectrum shape in redshift-space to the precision required for current and future surveys \citep{2021arXiv211010161I}. 

Application of bispectrum models to true data is hampered by the effects of survey geometry. In particular, this leads to a triple convolution of the true bispectrum with the survey window function; an effect which is costly to replicate in practice, especially if one must sample the model many thousands of times. Previous works \citep{1982ApJ...259..474F,2000ApJ...544..597S,2001ApJ...546..652S,2015MNRAS.451..539G,2017MNRAS.465.1757G,2018MNRAS.478.4500P,2019MNRAS.484.3713G} have avoided this problem by making certain assumptions, most commonly, that the action of the window can be reproduced by instead window-convolving the \textit{power spectrum} (noting that the tree-level bispectrum depends on two power spectra). However, this assumption is unwarranted. Firstly, it introduces four copies of the survey mask rather than three (two per power spectrum), and secondly, it does not account for the full geometric structure of the statistic. This has led to previous studies removing triangle configurations most strongly affected by the window (those involving large-scale (soft) $k$-modes), which are also those in which signatures of new physics (such as primordial non-Gaussianity) can be most easily seen. If we wish to perform a full bispectrum study, such assumptions must be avoided. One route is to project the bispectrum onto a new basis \citep{2019MNRAS.484..364S,2020MNRAS.497.1684S}; this allows the window functions to be straightforwardly applied, though comes at the cost of having to similarly project the theoretical model.

\begin{figure}[!t]
	\centering
	\includegraphics[width=0.6\textwidth]{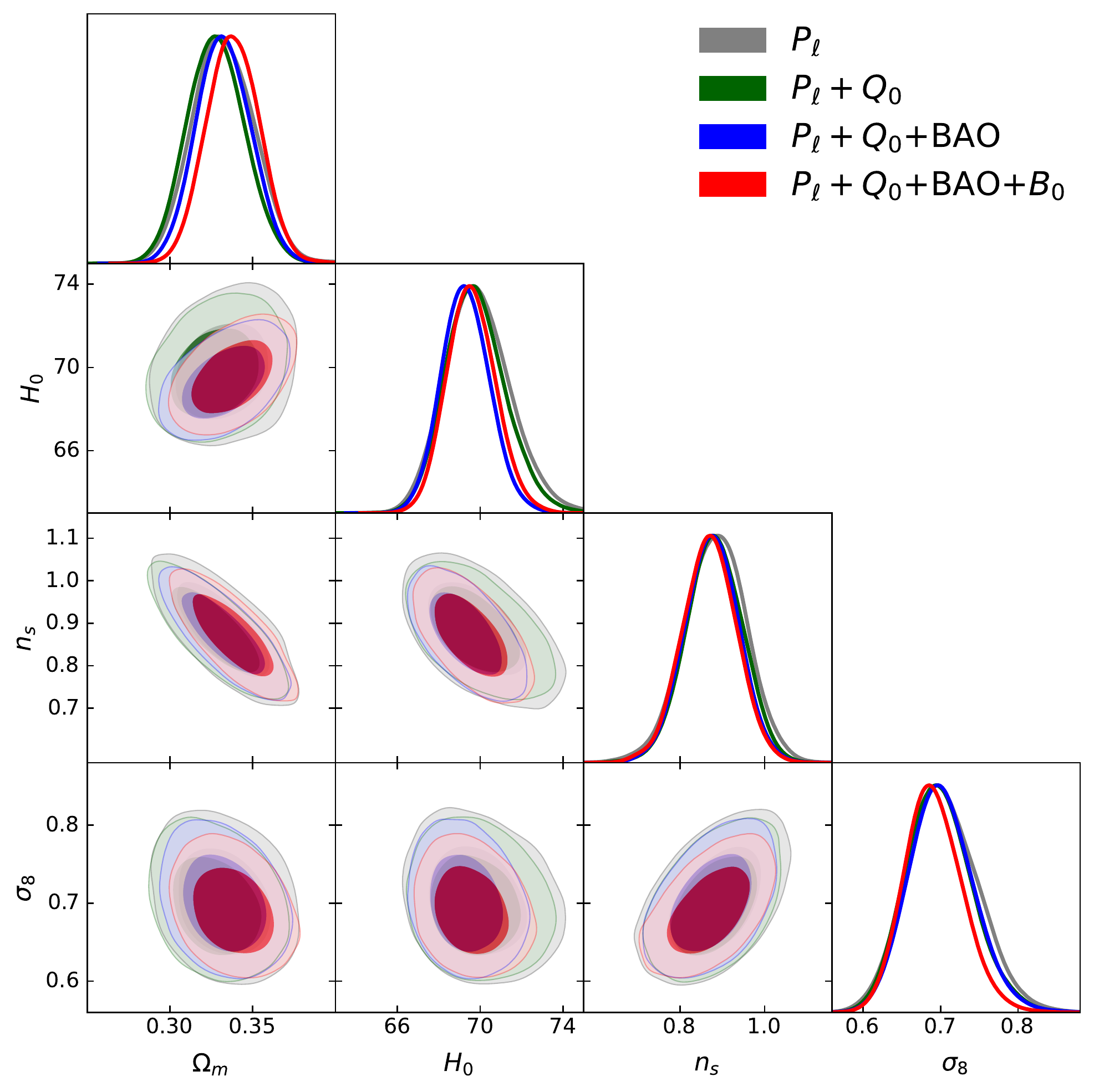}
	\caption{$\Lambda$CDM parameter constraints from the full BOSS DR12 galaxy dataset, using the following combinations of statistics, as well as a BBN prior on the baryon density: the (window-free) redshift-space power spectrum $P_\ell$ up to $k_{\rm max} = 0.2\hMpc$ (gray), the above plus the real-space power spectrum proxy $Q_0$ (green) up to $k_{\rm max} = 0.4\hMpc$ (green), the above plus BAO parameters from the post-reconstructed power spectrum (blue), the above plus the bispectrum monopole $B_0$ up to $k_{\rm max} = 0.08\hMpc$ (red). The bispectrum is the main new feature of this work, and leads to a tightening of $\sigma_8$ by $\approx 10\%$, with the modest improvement linked to our conservative analysis choices. The $\sigma_8$ posteriors are $\approx 1\sigma$ larger than those found in some former analyses; this is due to an error in the public power spectra, as discussed in \S\ref{subsec: comparison}. Tab.\,\ref{tab: main-results} gives the associated marginalized posteriors for each analysis shown above.}\label{fig: main-results}
\end{figure}

In this work, we make use of the power spectrum and bispectrum estimators proposed in \citep{2021PhRvD.103j3504P,2021arXiv210706287P}. Analogous to methods discussed in the late 1990s \citep[e.g.][]{1997PhRvD..55.5895T,2002MNRAS.335..887T,1998ApJ...499..555T,2005astro.ph..3604H}, these estimate the true (unwindowed) power spectrum and bispectrum directly, and thus obviate the need to window-convolve the theory model. A na\"ive implementation of such approaches is computationally prohibitive, in particular, due to the need to compute the covariance matrix between each pixel in the dataset, and the requirement to estimate a large-dimensional Fisher matrix, which acts to deconvolve the statistic. However, the approach can be made efficient using modern computational techniques. Given these measurements, we perform a full analysis of the large-scale power spectrum and bispectrum of the BOSS DR12 galaxy data \citep{2015ApJS..219...12A}, utilizing the latest theoretical models derived within the EFTofLSS, in particular those of the one-loop power spectrum \citep{2020PhRvD.102l3541N}, reconstructed power spectrum \citep{2020JCAP...05..032P}, and tree-level bispectrum \citep{Desjacques:2018pfv,2021arXiv211010161I}.\footnote{An important aspect of our theoretical model is an efficient IR resummation scheme for the redshift-space power spectrum and bispectrum, such as that derived within the time-sliced perturbation theory approach \cite{Blas:2015qsi,Blas:2016sfa,Ivanov:2018gjr,Vasudevan:2019ewf}.}
This stretches beyond previous studies in three key ways: (1) the window function is self-consistently treated, requiring no large-scale modes to be excluded, (2) we include a full treatment of all physical effects, (3) we measure cosmological parameters directly, instead of heuristic scaling parameters (such as BAO distortion parameters and amplitude ratios). In the current work, we consider only constraints on $\Lambda$CDM physics; however, an important extension is to measure non-standard parameters, in particular those of primordial non-Gaussianity (see \citep{2018MNRAS.478.1341K} for a recent forecast). This will be discussed in future work.

\begin{table}[!t]
    \centering
    \rowcolors{2}{white}{vlightgray}
  \begin{tabular}{|c|cccc|} \hline
    \textbf{Dataset} & $\Omega_m$ & $H_0$ & $n_s$ & $\sigma_8$\\\hline
   $P_\ell(k)$ & $\quad 0.332_{-0.020}^{+0.019}\quad $ & $\quad 69.9_{-1.7}^{+1.5}\quad $ & $0.883_{-0.072}^{+0.076}$ & $\quad 0.704_{-0.049}^{+0.044}\quad $\\
   $P_\ell(k)+Q_0(k)$ & $0.328_{-0.019}^{+0.017}$ & $69.8_{-1.6}^{+1.3}$ & $0.880_{-0.068}^{+0.068}$ & $0.699_{-0.046}^{+0.040}$\\
   $P_\ell(k)+Q_0(k)+\mathrm{BAO}$ & $0.333_{-0.018}^{+0.016}$ & $69.3_{-1.3}^{+1.1}$ & $0.874_{-0.064}^{+0.067}$ & $0.701_{-0.045}^{+0.040}$\\
  \quad$P_\ell(k)+Q_0(k)+\mathrm{BAO}+B_0$\quad\quad  & $0.338_{-0.017}^{+0.016}$ & $69.6_{-1.3}^{+1.1}$ & $0.870_{-0.064}^{+0.067}$ & $0.692_{-0.041}^{+0.035}$\\\hline
    \end{tabular}
    \caption{Mean and 68\% confidence intervals for cosmological parameters from the main analysis of this work (matching Fig.\,\ref{fig: main-results}). A full summary of these these analyses can be found in Tab.\,\ref{tab: free-ns}, with the analogous results from an analysis with \textit{Planck} priors on $n_s$ given in Tab.\,\ref{tab:fix-ns}. $H_0$ posteriors are given in $\hun$ units.}
\label{tab: main-results}
\end{table}

\vskip 4pt

The structure of this paper is as follows. 
Our main cosmological results are summarized in Fig.\,\ref{fig: main-results} and in Tab.\,\ref{tab: main-results}, with the power spectrum and bispectrum measurements shown in Fig.\,\ref{fig: pk-bk-plot}, and bias parameter constraints depicted in Fig.\,\ref{fig:bias}. We begin by describing the observational data in \S\ref{sec: data}, before discussing our power spectrum and bispectrum estimators in \S\ref{sec: estimators}. \S\ref{sec: model} presents our perturbative model for the power spectrum and bispectrum (including extension to the real-space power spectrum, to allow for a larger $k$-range), as well as the likelihood used in our analyses. The methodology is tested in \S\ref{sec: consistency-tests}, before we present the main results of our analysis in \S\ref{sec: results}. We conclude with a summary and discuss future work in \S\ref{sec: summary}, with additional parameter constraints given in Appendix \ref{appen: full-constraints}. 
\resub{Our measurements and likelihoods} are made publicly available online.\footnote{\resub{See \href{https://github.com/oliverphilcox/BOSS-Without-Windows}{GitHub.com/oliverphilcox/BOSS-Without-Windows} and \href{https://github.com/oliverphilcox/full\_shape\_likelihoods}{GitHub.com/oliverphilcox/full\_shape\_likelihoods}.}}

\section{Datasets}\label{sec: data}

The primary dataset of this work is the twelfth data-release (DR12) of the Baryon Oscillation Spectroscopic Survey (BOSS) \citep{2015ApJS..219...12A}, part of SDSS-III \citep{2011AJ....142...72E}. In total, the survey contains over a million galaxies, observed in two disjoint regions of the sky; the Northern and Southern galactic caps (NGC and SGC). We divide the survey into two contiguous redshift slices, hereafter `z1' and `z3', encompassing $0.2<z<0.5$ and $0.5<z<0.75$ respectively (with effective redshifts 0.38 and 0.61), giving a total of four chunks. This is the decomposition used in previous power spectrum analyses \citep[e.g.,][]{2017MNRAS.466.2242B,2020JCAP...05..042I,2020JCAP...05..005D} though we note that it mixes galaxies from the CMASS and LOWZ samples, which will have implications for the effective bias parameters. In full, we use the publicly available `CMASSLOWZTOT' galaxy samples,\footnote{Available at \href{https://data.sdss.org//sas/dr12/boss/lss}{data.sdss.org/sas/dr12/boss/lss}} filtered by redshift, and define the survey window function using the public \textsc{mangle} mask (required for the estimators of \S\ref{sec: estimators}) as well as a set of random particles, with fifty times the galaxy density. The latter are additionally used to define the overdensity field. As in previous works, we use the following galaxy weights:
\beq\label{eq: boss-weights}
    w_{\rm tot} = (w_{\rm rf}+w_{\rm fc}-1)w_{\rm sys},
\eeq
encoding redshift-failure, fiber-collision and systematic effects respectively \citep{2017MNRAS.466.2242B}.\footnote{Note that optimality weights (such as the FKP weight) are naturally included in the estimators of \S\ref{sec: estimators}.} 

To construct a Gaussian likelihood for our dataset, we require a covariance matrix, which is here obtained using a (publicly available) suite of 2048 `MultiDark-Patchy' mock catalogs (hereafter \textsc{Patchy}) \citep{2016MNRAS.456.4156K,2016MNRAS.460.1173R}. Each is generated using an approximate gravity solver, and calibrated to an $N$-body simulation, providing an approximate form for the galaxy survey. These utilize a similar survey mask to the BOSS data and are generated using the cosmology $\{\Omega_m = 0.307115, \sigma_8 = 0.8288, h = 0.6777, \sum m_\nu = 0\}$. A weight is associated with each particle:
\beq\label{eq: patchy-weights}
    w_{\rm tot} = w_{\rm veto}w_{\rm fc},
\eeq
involving a veto mask and a fiber collision term. We use a separate random catalog for the \textsc{Patchy} simulations.

Due to their approximate nature, the \textsc{Patchy} mocks are not sufficiently accurate to test our analysis pipeline. Instead, we make use of the 84 \textsc{Nseries} mock catalogs\footnote{Available at \href{https://www.ub.edu/bispectrum/page11.html}{www.ub.edu/bispectrum/page11.html}} \citep{2017MNRAS.470.2617A}; these are computed from full $N$-body simulations (though are not fully independent) using similar selection functions and halo occupation distribution to the BOSS data, as well as full treatment of fiber collisions. The \textsc{Nseries} mocks use the true cosmology $\{\Omega_m = 0.286, \sigma_8 = 0.82, n_s = 0.97, h = 0.7, \sum m_\nu = 0\}$. Unlike the BOSS data, galaxies are not split into `z1' and `z3' chunks; rather, the simulations include only the CMASS NGC sample, spanning $0.43<z<0.7$ (with effective redshift $z_{\rm eff}=0.56$). This has a different geometry to the BOSS data, but due to its large cumulative volume, provides a sensitive test of our methodology, in particular that of systematic and geometric effects. The dataset contains completeness weights and has an associated random catalog. Since the \textsc{Nseries} data uses a different survey geometry, it requires a different covariance matrix; here this is generated using 2048 \textsc{Patchy} simulations with the \textsc{Nseries} window function (hereafter dubbed `\textsc{Patchy-Nseries}`). In total, we have four different random catalogs - BOSS, \textsc{Patchy}, \textsc{Nseries} and \textsc{Patchy-Nseries} - each of which are separately treated using the estimators of \S\ref{sec: estimators}. In all cases, data are converted to Cartesian coordinates assuming the fiducial cosmology: $\{\Omega_{m,\rm fid} = 0.31, h_{\rm fid} = 0.676\}$, which is used to calibrate the geometric distortion parameters (\textit{i.e.}\ Alcock-Paczynski parameters) of \S\ref{sec: model}.

Finally, we include a prior on the physical baryon density of $\omega_b=0.02268\pm 0.0038$, from Big Bang Nucleosynthesis (BBN) considerations \citep{2015JCAP...07..011A,2018ApJ...855..102C,2019JCAP...10..029S}; this has a similar width to that from \textit{Planck} \citep{2020A&A...641A...6P}. Whilst we keep the tilt, $n_s$, free in our baseline analyses, we additionally consider a scenario in which it is fixed to the \textit{Planck} best-fit ($n_s = 0.9649$), to allow for comparison with other full-shape analyses.

\section{Estimators}\label{sec: estimators}

Previous works have constrained cosmology from the power spectrum and bispectrum by measuring the statistics using simple FKP-like estimators \citep[e.g.][]{1994ApJ...426...23F,2015PhRvD..92h3532S,2015MNRAS.451..539G,2015MNRAS.453L..11B,2017MNRAS.466.2242B}. Since the true galaxy distribution is modulated by a survey mask, the spectra are convolved with a non-trivial function of this mask, which must be replicated in the theoretical model. Whilst this can be simply implemented using FFTs for the power spectrum, it is more difficult for the bispectrum, since one must perform a six-dimensional convolution integral. This has led to previous studies excising regions of the spectrum most affected by the window, leading to a loss of information. Here, we adopt a different approach, instead measuring the unwindowed power spectrum and bispectrum directly from the observational data.

To do this, we use the quadratic and cubic estimators discussed in  \citep{2021PhRvD.103j3504P} and \citep{2021arXiv210706287P}. These are derived by first writing down the large-scale likelihood for the galaxy survey and analytically optimizing for the statistics of interest, in a manner similar to that used in early CMB and LSS analyses \citep[e.g.][]{1997PhRvD..55.5895T,2002MNRAS.335..887T,1998ApJ...499..555T,2005astro.ph..3604H}. Schematically, the estimator for the power spectrum in some bin $\alpha$ (comprising a set of wavenumbers and multipole) can be written:
\beq
    \hat p_\alpha = \sum_{\beta}F_{\alpha\beta}^{-1}\left[\hat q_\beta - \bar q_\beta\right],
\eeq
where $\hat q_\alpha$ is a quadratic piece, $\bar q_\alpha$ is a bias term, and $F_{\alpha\beta}$ is a Fisher matrix. These can be computed in terms of the pixelized data-vector, $\vec d$, the associated covariance matrix, $\C\equiv\av{\vd\vd^T}$, and the noise covariance, $\N$, as
\beq
    \hat q_\alpha = \frac{1}{2}\left[\Ci\vd\right]^T \C_{,\alpha}\left[\Ci\vd\right], \qquad \bar q_\alpha = \frac{1}{2}\Tr{\Ci\C_{,\alpha}\Ci\N}, \qquad F_{\alpha\beta} = \frac{1}{2}\Tr{\Ci\C_{,\alpha}\Ci\C_{,\beta}},
\eeq
where $\C_{,\alpha}$ represents the derivative of the pixel covariance with respect to the signal of interest, and we note that both $\C$ and $\N$ both depend on the survey mask. In certain limits, the first piece is proportional to the conventional FKP power spectrum, whilst the second removes the Poissonian noise contribution, and the third acts to deconvolve the window. A similar form can be obtained for the bispectrum components, $\{b_\alpha\}$ (where $\alpha$ specifies a triplet of bins):
\beq
    \hat b_\alpha = \sum_{\beta}\mathcal{F}^{-1}_{\alpha\beta}\hat c_\beta, 
\eeq
where $\hat c_\alpha$ is a cubic estimator and $\mathcal{F}$ is the associated Fisher matrix \resub{(which is close to tridiagonal)}. These can be written in terms of the three-point expectation of the data: $\B^{ijk} \equiv \av{d^id^jd^k}$ (using latin indices to denote pixels and employing Einstein summation), and take the form
\beq
    \hat c_\alpha = \frac{1}{6}\B^{ijk}_{,\alpha}\left[\Ci\vec d\right]_i\left(\left[\Ci\vec d\right]_j\left[\Ci\vec d\right]_k-3\C_{jk}^{-1}\right), \qquad \mathcal{F}_{\alpha\beta} = \frac{1}{6}\B_{,\alpha}^{ijk}\B_{,\beta}^{lmn}\Ci_{il}\Ci_{jm}\Ci_{kn}.
\eeq
Notably, $\hat c_\alpha$ contains a piece \textit{linear} in the data; this is not found in conventional LSS estimators (though commonplace for the CMB \citep{2011MNRAS.417....2S}), and helps to reduce variance on large scales. All quantities appearing in the power spectrum and bispectrum estimators can be efficiently computed using Monte Carlo methods, and give minimum variance estimates of the spectra. 


Here, we use the specific forms of the estimators suggested in \citep{2021arXiv210706287P}, approximating the pixel covariance by a diagonal (FKP-like) form with $P_{\rm fkp}\approx 10^4h^{-3}\mathrm{Mpc}^3$, though still fully accounting for survey geometry.\footnote{Due to a minor error in the original code, the FKP weight differs from this value at the $\sim 10\%$ level; the impact of this is negligible in practice.} Whilst this leads to a slight loss of optimality, this approximation greatly reduces computation time and does not induce bias. The estimators are implemented using a Fourier-space grid with Nyquist frequency of $k_{\rm Nyq} = 0.45\hMpc$ ($0.3\hMpc$) for the power spectrum (bispectrum), measuring all modes up to $k = 0.41\hMpc$ ($0.16\hMpc$) with $\Delta k = 0.005\hMpc$ ($0.01\hMpc$), and using the power spectrum monopole, quadrupole and hexadecapole. We caution that the first and last bins will be contaminated with integral-constraint and window-leakage effects respectively; these are not included in the analysis below. In total we use 100 Monte Carlo simulations to compute the Fisher matrices (which gives only a subpercent noise penalty, cf.\,\citep{2021arXiv210706287P}), requiring a total of $\approx 10\,000$ ($\approx 3000$) CPU-hours for the power spectrum (bispectrum) of all four BOSS data chunks. Importantly, this only needs to be done once per survey geometry. The increased computation time for the power spectrum is due to both the larger number of bins (the computation time scales as $N_{\rm bins}$) and the higher-resolution grid. The data-dependent part of the power spectrum (bispectrum) estimator requires $\approx 15$ ($\approx 10$) CPU-minutes to compute per simulation, and thus can be easily applied to all $2048$ \textsc{Patchy} mocks.

\begin{figure}
    \centering
    \includegraphics[width=\textwidth]{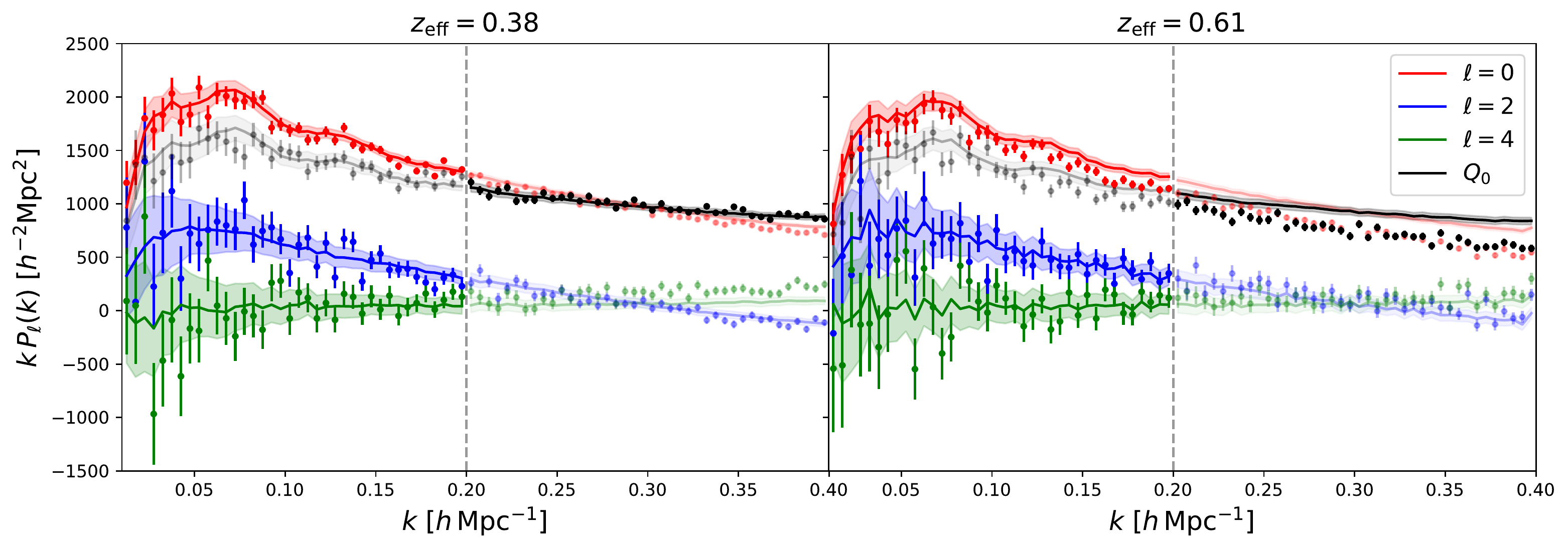}
    \includegraphics[width=\textwidth]{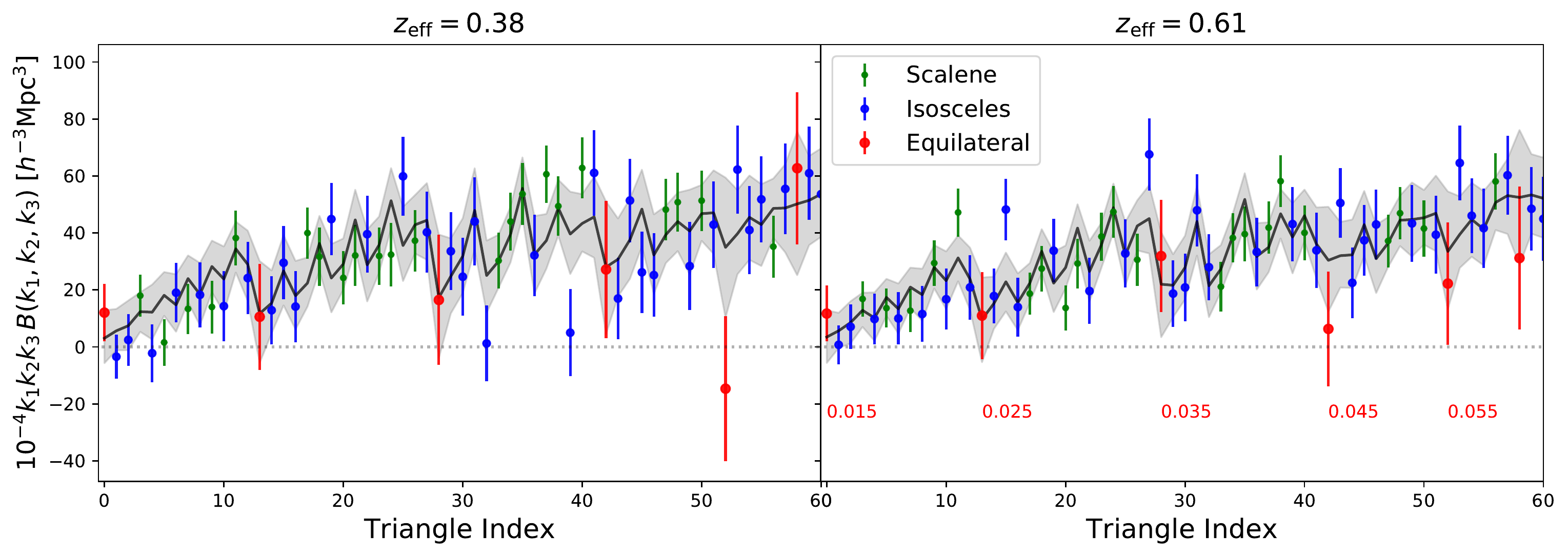}
    \caption{Measured power spectra (top) and bispectra (bottom) from the BOSS dataset (points) and 2048 \textsc{Patchy} mocks (lines and shaded regions) for two redshift bins `z1' (left) and `z3' (right). For the power spectra, we show measurements from the monopole, quadrupole, and hexadecapole, in red, blue, and green respectively, as well as the $Q_0$ statistic, $Q_0(k)\equiv P_0(k)-(1/2)P_2(k)+(3/8)P_4(k)$, which is a proxy for the real-space power spectrum. The vertical line disambiguates regions fit with the full power spectrum multipoles (left) and those with $Q_0$; the other regions (shown in faint lines) are not used in the analysis. For bispectra, we plot all triangle bins included in the analysis with $k<0.08\hMpc$, noting that the observed structure arises from the bin ordering. These are ordered by triangle side, with scalene, isosceles, and equilateral triangles shown in green, blue, and red respectively. The red numbers in the right panel give the value of $k$ for each equilateral bin. For clarity, we have combined estimates from the NGC and SGC regions (weighting by their sky fractions, with $f_{\rm NGC}\approx 0.7$); these are treated as separate samples in the main analysis of this work.}
    \label{fig: pk-bk-plot}
\end{figure}

Using the output of the above estimators, we form the following datasets, which will be analyzed in \S\ref{sec: results} (cf.\,Fig.\,\ref{fig: main-results}):
\begin{itemize}
    \item \textbf{Redshift-Space Power Spectrum, $P_\ell(k)$}: Here, we use the power spectrum monopole, quadrupole and hexadecapole (with $\ell\in\{0,2,4\}$) with $k\in[0.01,0.20]\hMpc$ and $\Delta k = 0.005\hMpc$. Our $k$-space cuts are motivated by the results of \citep{2020PhRvD.102l3541N,2021PhRvD.103b3507C}. These are corrected for the effects of survey geometry and given by the outputs of the above estimators. We use a total of $38$ bins for each multipole.
    \item \textbf{Real-Space Power Spectrum, $Q_0(k)$}: Following \citep{2021arXiv211000006I} (see also the earlier works of \cite{Hamilton:2000du,Scoccimarro:2004tg}), we use an analog to the real-space power spectrum computed from the above redshift-space power spectrum multipoles up to $\ell=4$, explicitly via $Q_0(k) \equiv P_0(k)-\tfrac{1}{2}P_2(k)+\tfrac{3}{8}P_4(k)$. This is included for $k\in[0.2,0.4]\hMpc$ (involving $40$ bins) and allows cosmological information to be included without the limitations of fingers-of-God modeling. 
    \item \textbf{BAO Parameters, $\alpha_{\parallel}, \alpha_\perp$}: To capture sound-horizon information present in the reconstructed power spectrum, we include the parallel and perpendicular BAO scaling parameters, as discussed in \citep{2020JCAP...05..032P}. These are measured for all data chunks in the BOSS and \textsc{Patchy} datasets, and correlations are included via a joint covariance matrix (see \citep{2021arXiv211005530C} for an alternative approach). We note that only 999 mocks are available for these parameters.
    \item \textbf{Bispectrum, $B_0(k_1,k_2,k_3)$}: The bispectrum monopole is obtained from the above cubic estimators.\footnote{Strictly, the estimate of the bispectrum monopole is unbiased only if one measures all possible bispectrum multipoles, due to coupling of the redshift-space contributions with the anisotropic window function (analogous to the power spectrum case). Here, we assume this effect to be small, though its magnitude will be studied in future work.} To avoid window leakage and poorly modelled regimes, we use only $k$-modes in the range $[0.01,0.08]\hMpc$ and $\Delta k = 0.01\hMpc$, whose centers satisfy the triangle condition $|k_1-k_2|\leq k_3\leq k_1+k_2$ \citep{2021arXiv211010161I}.\footnote{\resub{Note that triangles violating this condition must be included in the initial bispectrum estimator, to avoid bias, though they can be dropped when the statistic is analyzed.}} In total, we include $62$ bispectrum bins.
\end{itemize}
Each statistic is computed for the BOSS data, the \textsc{Patchy} mocks, the \textsc{Nseries} mocks and the \textsc{Patchy-Nseries} mocks, and the data, \resub{estimation pipeline, and likelihoods} are made publicly available online.\footnote{\resub{See \href{https://github.com/oliverphilcox/BOSS-Without-Windows}{GitHub.com/oliverphilcox/BOSS-Without-Windows} and \href{https://github.com/oliverphilcox/full\_shape\_likelihoods}{GitHub.com/oliverphilcox/full\_shape\_likelihoods}.}} The total data-vector contains $N_{\rm bin}=114+40+2+62=218$ elements, and is shown in Fig.\,\ref{fig: pk-bk-plot} (cf.\,\S\ref{sec: results}).

\section{Theoretical Model and Likelihood}\label{sec: model}

\subsection{Theory Model}
To model the window-free power spectrum and bispectrum, we utilize the EFTofLSS, as implemented in the Boltzmann code \textsc{class-pt} \citep{2020PhRvD.102f3533C} (see also \citep{2020JCAP...07..062C,2021JCAP...01..006D}). For consistency, the power spectrum (bispectrum) model is computed up to one-loop (tree-level) order, and both include full treatment of all necessary components, including perturbative corrections, galaxy bias, ultraviolet counterterms (to consistently account for short-scale physics), infrared resummation (to treat long-wavelength displacements) and stochasticity (including shot-noise and fingers-of-God effects). Full discussion of our models can be found in \citep{2020JCAP...05..042I,2020PhRvD.102f3533C,2020PhRvD.102l3541N} for the redshift-space power spectrum, \citep{2021arXiv211000006I} for the real-space power spectrum analog, \citep{2020JCAP...05..032P} for the BAO parameters, and \citep{2021arXiv211010161I} for the bispectrum.  

Schematically, our model of the power spectrum multipoles takes the following form (before the effects of infrared resummation and coordinate transformations):
\beq\label{eq: power-spectrum-model}
	P_\ell(k) = P_\ell^{\rm tree}(k) + P_{\ell}^{\rm 1-loop}(k) + P_{\ell}^{\rm ct}(k) + P_\ell^{\rm stoch}(k),
\eeq
where the four terms are the usual linear theory Kaiser multipoles (scaling as the linear-theory power spectrum, $P_{\rm lin}(k)$), the one-loop perturbation theory corrections (scaling as $k^2P_{\rm lin}(k)$ on large scales), the counterterms (scaling as $k^2P_{\rm lin}(k)$), and the stochastic contributions (scaling as a constant, plus corrections), respectively. This is then resummed to correct for the action of non-perturbative long-wavelength displacements, with the effect of suppressing the wiggly part of the spectrum (see \citep{2020PhRvD.102f3533C}). We account for the effects of an incorrect fiducial cosmology (often known, erroneously, as the Alcock-Paczynski distortion \citep{1979Natur.281..358A}) via the rescalings
\beq\label{eq: coord-rescaling}
	k \to k' &\equiv& k\left[\left(\frac{H_{\rm true}}{H_{\rm fid}}\right)^2\mu^2+\left(\frac{D_{A,\rm fid}}{D_{A,\rm true}}\right)^2(1-\mu^2)\right]^{1/2}\\\nonumber
	\mu \to\mu' &\equiv& \mu\left(\frac{H_{\rm true}}{H_{\rm fid}}\right)\left[\left(\frac{H_{\rm true}}{H_{\rm fid}}\right)^2\mu^2+\left(\frac{D_{A,\rm fid}}{D_{A,\rm true}}\right)^2(1-\mu^2)\right]^{-1/2},
\eeq
where unprimed quantities are those measured observationally, and all quantities are evaluated at the sample redshift. The real-space power spectrum model (needed for the $Q_0$ statistic) is similar, but obtained by summing over the redshift-space multipoles entering \eqref{eq: power-spectrum-model}, via $Q_0(k) = P_0(k)-\tfrac{1}{2}P_2(k)+\tfrac{3}{8}P_4(k)$.\footnote{Note that this is not quite equal to the real-space power spectrum (\textit{i.e.}\, that with $f=0$) due to infrared resummation effects.}

The bispectrum model can be written schematically in 3D space as
\beq\label{eq: bispectrum-model}
	B(\vk_1,\vk_2) = B^{\rm tree}(\vk_1,\vk_2) + B^{\rm ct}(\vk_1,\vk_2) + B^{\rm stoch}(\vk_1,\vk_2),
\eeq
corresponding to the tree-level perturbative contributions, counterterms and stochasticity respectively. Whilst counterterms strictly appear only at one-loop, \citep{2021arXiv211010161I} found it necessary to include a $k^2\mu^2 \equiv k_{\parallel}^2$-like correction in the tree-level model to encapsulate fingers-of-God effects. The three terms scale as $P_{\rm lin}^2(k)$, $k_{\parallel}^2P_{\rm lin}^2(k)$ and a constant plus $P_{\rm lin}(k)$. We include the effects of infrared resummation by replacing the linear matter power spectrum by its resummed version (with wiggles suppressed), and the effects of the fiducial cosmology via a redefinition of wavenumbers and angles entering \eqref{eq: bispectrum-model}, similar to \eqref{eq: coord-rescaling}. The model is then integrated over external angles to obtain the bispectrum monopole, $B_0(k_1,k_2,k_3)$, which can be directly compared to data without additional window convolutions. Note that we also include a multiplicative discreteness weight, to account for the finite resolution of the Fourier-space grid \citep{2021arXiv211010161I}.

\subsection{EFTofLSS Parameters}\label{subsec: nuisance-parameters}
The full model for the power spectrum and bispectrum is specified by the following nuisance parameters, which appear in one or both statistics (usings the conventions described in \citep{2021arXiv211010161I}):
\beq\label{eq: nuisance-params}
	\{b_1,b_2,b_{\mathcal{G}_2}, b_{\Gamma_3}\} \times \{c_{0}, c_{2}, c_{4}, \tilde c, c_1\} \times \{P_{\rm shot}, a_0, a_2, B_{\rm shot}\}
\eeq
where the first set encodes galaxy bias (from linear, quadratic, tidal, and third-order effects respectively), the second give the counterterms for the monopole, quadrupole, and hexadecapole, fingers-of-God effect and bispectrum, whilst the final set accounts the stochastic nature of the density field. Since the BOSS regions have different selection functions and calibrations, we allow the parameters to vary freely in each of the four data chunks, giving a total of 42 free parameters. 

In the likelihood described below, we assume the following bias parameter priors, following \citep{2021arXiv211010161I}:
\be
\begin{split}
& b_1\in \text{flat}[0,4]\,, \quad b_2\sim \mathcal{N}(0,1^2)\,, 
\quad b_{\mathcal{G}_2}\sim \mathcal{N}(0,1^2) \,,
\quad
b_{\Gamma_3}\sim \mathcal{N}\left(\frac{23}{42}(b_1-1),1^2\right),
\end{split}
\ee
where $\mathcal{N}(\mu,\sigma^2)$ indicates a Gaussian distribution with mean $\mu$ and variance $\sigma^2$. Similarly, we use Gaussian priors for the counterterms and stochasticity parameters:
\be
\begin{split}
& \frac{c_0}{[\text{Mpc}/h]^2} \sim \mathcal{N}(0,30^2)\,,\quad 
\frac{c_2}{[\text{Mpc}/h]^2} \sim \mathcal{N}(30,30^2)\,,\quad \frac{c_4}{[\text{Mpc}/h]^2} \sim \mathcal{N}(0,30^2)\,,\quad 
\frac{\tilde{c}}{[\text{Mpc}/h]^4} \sim \mathcal{N}(500,500^2)\,,\\
& \frac{c_1}{[\text{Mpc}/h]^2} \sim \mathcal{N}(0,50^2)\,,\quad P_{\rm shot} \sim \mathcal{N}(0,2^2)\,,\quad a_{0}
\sim \mathcal{N}(0,2^2)\,,\quad a_2\sim \mathcal{N}(0,2^2)\,,\quad 
B_{\rm shot}\sim \mathcal{N}(1,2^2).
\end{split}
\ee
In our conventions, the stochastic contributions are specified by the scale-dependent spectra, $P_{\rm stoch}$ and $B_{\rm stoch}$, with
\beq
    P_{\rm stoch}(k)&=&\frac{1}{\bar n}\left[1+P_{\rm shot} + a_0 \left(\frac{k}{k_{\rm NL}}\right)^2 + a_2\mu^2 \left(\frac{k}{k_{\rm NL}}\right)^2 \right],\\\nonumber 
    B_{\rm stoch}(k_1,k_2,k_3) &=& \frac{B_{\rm shot}}{\bar n}\left[Z_1({\bf k}_1)(b_1+b_\eta f\mu^2_1)P^{\rm tree}(k_1)+\text{cycl.}\right] + \frac{1}{\bar n^2}\left(1+P_{\rm shot}\right)^2,
\eeq\label{eq:Pstoch}
for non-linear scale $k_{\rm NL}=0.45~\hMpc$, redshift-space kernel $Z_1(\vk)$, number density $\bar n\approx 3\times 10^{-4}\,h^3\,\mathrm{Mpc}^{-3}$, and $b_\eta =(1+P_{\rm shot})/(B_{\rm shot})$.

\subsection{Likelihood}
Given that our analysis primarily concerns large-scale density modes, we will assume a Gaussian likelihood for the data-vector $\vec d$ (comprising the power spectrum multipoles, the real-space power spectrum $Q_0$, the BAO parameters, and the bispectrum), given some set of parameters $\vp$:
\beq\label{eq: likelihood}
	-2\log\mathcal{L}\left(\vec d|\vec p\right) &=& \left(\vec m(\vec p)-\vec d\right)^T\mathsf{C}^{-1} \left(\vec m(\vec p)-\vec d\right) + \log\left|2\pi\mathsf{C}\right|,
\eeq
where $\vec m(\vec p)$ is the model prediction. The covariance matrix $\mathsf{C}$ (including all cross-correlations) is computed using a set of $N_{\rm mocks} = 2048$ \textsc{Patchy} simulations:
\beq
	\mathsf{C} = \frac{1}{N_{\rm mocks}-1}\sum_{k=1}^{N_{\rm mocks}}\left(\vec d^k-\bar{\vec d}\right)\left(\vec d^k-\bar{\vec d}\right)^T,
\eeq
where $\vec d^k$ is the (stacked) data-vector from the $k$-th mock, and $\bar{\vec d}$ the average over mocks. We include also a Hartlap correction factor to debias the estimate of the inverse covariance \citep{2007A&A...464..399H}, setting $\mathsf{C}^{-1}\to\left(N_{\rm mocks}-N_{\rm bins}-2\right)/\left(N_{\rm mocks}-1\right)\times\mathsf{C}^{-1}$.\footnote{This correction is strictly an approximation; properly one should account for noise in the covariance matrix by replacing the likelihood of \eqref{eq: likelihood} with the modified $t$-distribution of \citep{2016MNRAS.456L.132S}. Here, we are in the limit of $N_{\rm mocks}\gg N_{\rm bins}$, thus thus has little effect, and we do not include it since it complicates the analytic marginalization over nuisance parameters.} When the BAO parameters are included in the analysis (obtained from reconstructed power spectrum data), $N_{\rm mocks}$ is reduced to $999$, slightly increasing the Hartlap factor. An alternative approach would be to compute the covariance matrix analytically \cite{2020MNRAS.497.1684S,Wadekar:2019rdu}; this naturally avoids sampling noise.

In the analyses below, we vary the following cosmological parameters:\footnote{Strictly speaking, we also fix the CMB temperature monopole $T_0$ to the FIRAS best-fit value. The effect of $T_0$ on cosmological observables has been studied in \citep{Ivanov:2020mfr}.}
\beq
	\{h, \resub{\ln(10^{10}A_s)}, \omega_{\rm cdm}, \omega_b, n_s\},
\eeq
from which we can obtain the derived parameters $H_0$, $\Omega_m$ and $\sigma_8$. Following \textit{Planck}, we fix the neutrino mass to $ \sum m_\nu=0.06\,\mathrm{eV}$ when analyzing BOSS data (or zero for \textsc{Patchy} simulations, matching the true value); this is done only for simplicity, and a full discussion of the impact of the bispectrum on neutrino mass measurement will be presented in future work. To convert the likelihood of \eqref{eq: likelihood} to a posterior, we require priors for all parameters: here we adopt flat priors on the cosmological parameters with infinite support, and use the priors given in \S\ref{subsec: nuisance-parameters} for the bias, counterterm and stochasticity parameters. Parameters that enter the theoretical model linearly, such as $P_{\rm shot}$, are marginalized over analytically, as in \citep{2021PhRvD.103d3508P}; this reduces the dimensionality and allows for faster sampling. To compute the parameter posteriors, we employ a Markov Chain Monte Carlo (MCMC) analysis, implemented within the \textsc{Montepython} code \citep{2019PDU....24..260B}. Sampling is performed across a number of parallel chains and assumed to converge when the Gelman-Rubin diagnostic satisfies $R<0.03$ for all parameters. Typically, this requires $\approx 500$ CPU-hours.

\section{Consistency Tests}\label{sec: consistency-tests}
Before presenting the main results of this work, we first test our pipeline by applying the estimators of \S\ref{sec: estimators} to the \textsc{Nseries} simulations discussed in \S\ref{sec: data}. In particular, we perform a full parameter inference study using likelihood described in \S\ref{sec: model}, with the data-vector given by the mean of 84 pseudo-independent \textsc{Nseries} mocks, including the power spectrum multipoles, the real-space power spectrum proxy, and the bispectrum monopole. As in \S\ref{sec: data}, we use a covariance matrix formed from 2048 \textsc{Patchy-Nseries} mocks, whose window function matches that of the CMASS NGC region. 

Two analyses are performed: (1), with the covariance rescaled by a factor of $84$ to give an effective volume equivalent to that of the total \textsc{Nseries} suite ($V_{\rm eff}=235~h^{-3}$Gpc$^3$), and (2), with the covariance rescaled by a factor of $2.4$, matching the total effective volume of BOSS DR12 ($V_{\rm eff}=6~h^{-3}\mathrm{Gpc}^3$). The first allows us to quantify the systematic errors given our choice of scale-cuts, whilst the second helps to assess the prior volume shifts arising from nuisance parameter marginalizations. We caution that the two sources of bias are very different in nature: the theoretical systematics are generated by truncating the perturbative expansion at a given order, whilst the prior volume effect appears due to the need to marginalize nuisance parameters over poorly known priors. In the limit of infinite data, \textit{i.e.}\ when nuisance parameters are determined by the data itself and not by the priors, the best-fit cosmological parameters will equal their true values if systematic effects can be ignored. We caution also that non-Gaussianity of the posterior can lead to mean values shifted with respect to the best-fits, even in the absence of priors and systematic errors. 



\begin{figure*}[!h]
\centering
\includegraphics[width=0.95\textwidth]{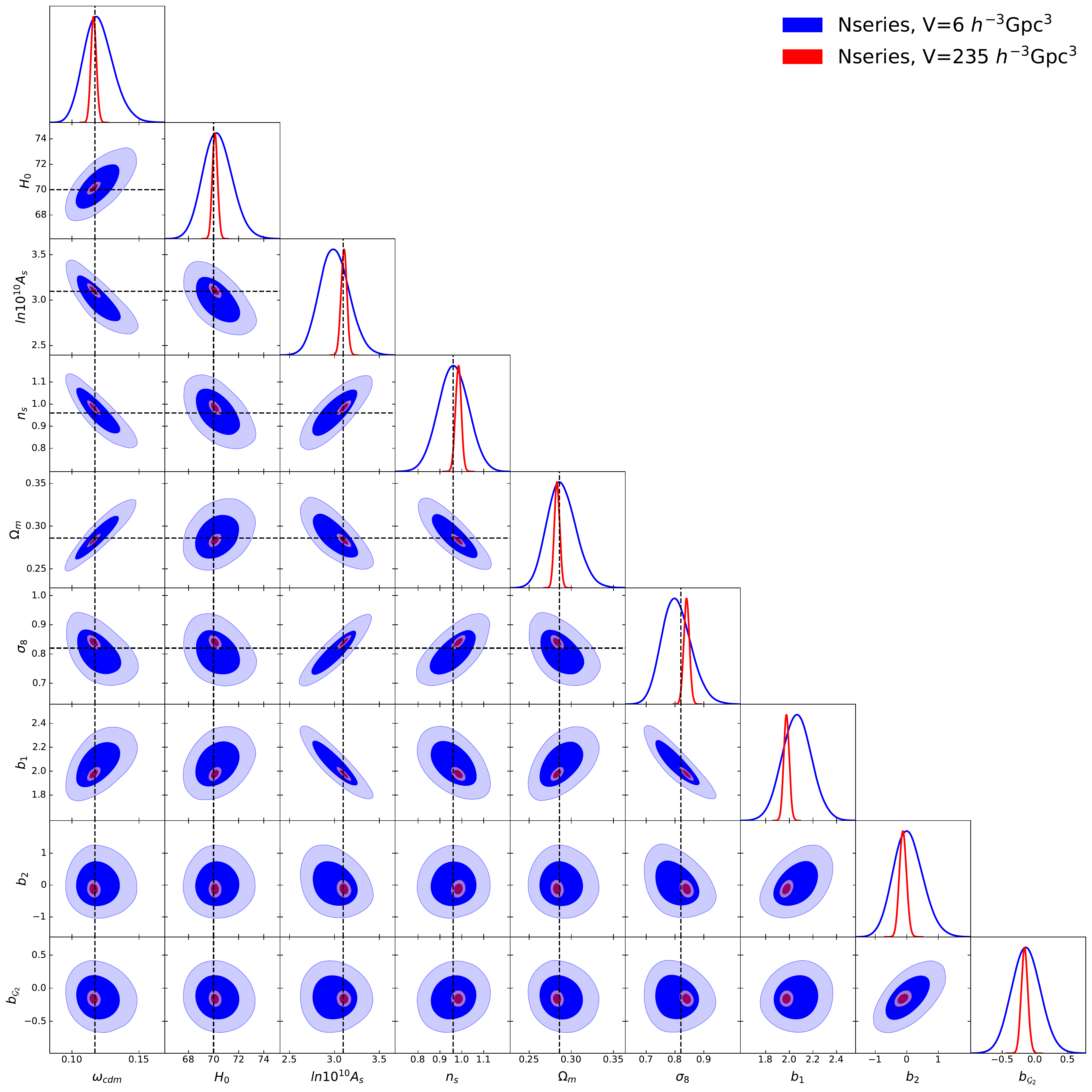}
\caption{Constraints on cosmological and bias parameters extracted from the power spectrum and bispectrum of 84 \textsc{Nseries} simulations, with a covariance matrix rescaled to match the total volume of BOSS (blue), and that of the \textsc{Nseries} suite (red). Dashed lines mark fiducial values of cosmological parameters, and we give the marginalized limits in Tab.\,\ref{tab:nseries}. Notably, any systematic effects are strongly subdominant for the BOSS-scaled posteriors, though there are slight shifts for the full \textsc{Nseries} volume (which is somewhat larger than the full DESI dataset).}\label{fig:Nseriescorner}  
\end{figure*}

\begin{table}[!t]
    \centering
    \rowcolors{2}{white}{vlightgray}
  \begin{tabular}{|c|ccccccccc|} \hline
   \textbf{Volume} 
   &  $\omega_{\rm cdm}$
   & $H_0$
   & $n_s$
   & $\ln(10^{10}A_s)$
   &  $\Omega_m$
   & $\sigma_8$
   & $b_1$
   & $b_2$
   & $b_{\mathcal{G}_2}$
      \\ [0.1cm]
\hline
$6\,h^{-3}\mathrm{Gpc}^3$
& $0.119_{-0.012}^{+0.010}$
& $70.3_{-1.3}^{+1.1}$
& $0.965_{-0.071}^{+0.068}$
& $3.01_{-0.17}^{+0.16}$
& $0.288_{-0.018}^{+0.016}$
& $0.808_{-0.054}^{+0.047}$
& $2.06_{-0.13}^{+0.12}$
& $0.04_{-0.52}^{+0.43}$
& $-0.14_{-0.23}^{+0.22}$
\\ 
\,$235\,h^{-3}\mathrm{Gpc}^3$\,
&  \,$0.1161_{-0.0022}^{+0.0020}$\,
 & \,$70.11_{-0.22}^{+0.20}$\,
  & \,$0.984_{-0.013}^{+0.014}$\,
  & \,$3.105_{-0.032}^{+0.033}$\,
  & \,$0.2831_{-0.0033}^{+0.0032}$\, 
  & \,$0.84_{-0.01}^{+0.01}$\,
  & $1.98_{-0.024}^{+0.023}$\,
  & \,$-0.12_{-0.12}^{+0.11}$\,
  & \,$-0.16_{-0.052}^{+0.050}$\,
\\  
\hline
\end{tabular}
\caption{Marginalized constraints on key cosmological and bias parameters extracted from the mean of 84 \textsc{Nseries} simulations, analyzed covariances corresponding to the BOSS volume (top) and the total 
\textsc{Nseries} volume (bottom). In all cases, the analysis uses the power spectrum multipoles up to $k_{\rm max} = 0.2\hMpc$, the real-space power spectrum up to $k_{\rm max} = 0.4\hMpc$, and the bispectrum up to $k_{\rm max} = 0.08\hMpc$. The fiducial cosmology is given by $\omega_{\rm cdm}=0.1171$, $H_0=70\,\mathrm{km}\,\mathrm{s}^{-1}\mathrm{Mpc}^{-1}$, $n_s=0.96$, $\ln(10^{10}A_s)=3.091$, $\Omega_m = 0.286$, and $\sigma_8=0.82$. The two-dimensional posterior is shown in Fig.\,\ref{fig:Nseriescorner}.}
\label{tab:nseries}
\end{table}

A corner plot of the resulting parameter constraints is shown in Fig.~\ref{fig:Nseriescorner}, with one-dimensional marginalized limits given in Tab.\,\ref{tab:nseries}. When considering the BOSS-scaled covarinace, we find that all cosmological parameters are correctly recovered within 68\% confidence level (CL); indeed, the parameter shifts are much smaller than $1\sigma$ in most cases, indicating that our joint power spectrum and bispectrum analysis is producing accurate parameter estimates. Considering the full simulation volume, we find that all parameters are recovered within the 95\% CL, with largest discrepancies found for $n_s$ and $\sigma_8$. The residual shifts can be conservatively attributed to systematic limitations of our theoretical model, arising from higher-order perturbative terms, or small inaccuracies in our window function treatment. We caution that the \textsc{Nseries} volume is larger even than the full DESI sample, and further, that the $84$ \textsc{Nseries} mocks are not fully independent (since they are generated from the same $N$-body simulations), thus the above shifts are necessarily an overestimate. 

The largest systematic shift is observed for $\sigma_8$, which, if one rescales the full \textsc{Nseries}-volume constraint to the BOSS volume, reaches the level of $0.4\sigma$. Whilst not insignificant, this is likely inflated by (a) the non-independence of the \textsc{Nseries} mocks, and (b), the non-Gaussianity of the posterior surface, which will drive the mean away from the best-fit value. We may also consider the marginalization bias, \textit{i.e.}\ the difference between the means in the two analyses; this is equal to $0.6\sigma$ for $\sigma_8$ (using BOSS error-bars) and less than $0.3\sigma$ for other parameters. These two biases cancel each other for $\sigma_8$, such that the resulting shift is only $0.2\sigma$. These results agree with previous studies \citep{2021PhRvD.103b3507C,2021arXiv211010161I}, and allow us to conclude that our analysis pipeline, with the chosen $k$-space cuts, is robust and may be applied to BOSS data to yield accurate parameter constraints. Given that the \textsc{Nseries} volume is several times larger than that of DESI, and the combined covariance is an underestimate (due to the independence considerations), we expect that the current analysis could be similarly applied to future surveys.

\section{Parameter Constraints from BOSS}\label{sec: results}

Below, we present the key results of this work: parameter constraints from the full BOSS DR12 dataset, including the power spectrum (both pre- and post-reconstruction), and the bispectrum. The observational data used in this work is shown in Fig.\,\ref{fig: pk-bk-plot} for the two redshift-slices for both BOSS and \textsc{Patchy}; whilst the large-scale power spectrum and bispectrum are generally in agreement, we note differences between the mocks and data at larger-$k$, particularly for the `z3' power spectrum. This matches that found in previous works \citep{2020JCAP...05..042I,2017MNRAS.465.1757G}, and is a consequence of the lack of small-scale matching, and simplified halo physics treatments in the \textsc{Patchy} mocks.

\begin{figure}
    \centering
    \includegraphics[width=0.8\textwidth]{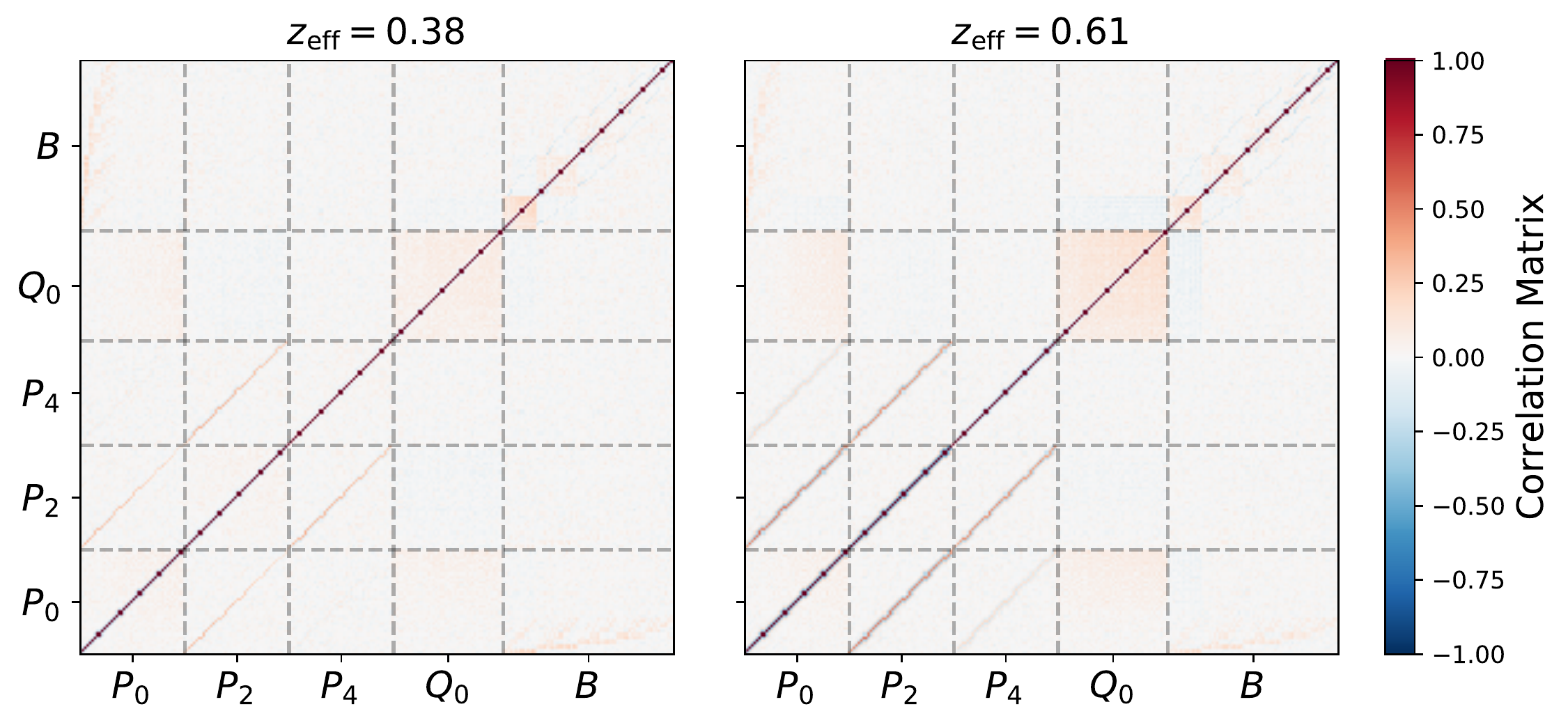}
    \caption{Correlation matrices for the power spectrum and bispectrum measurements used in this work, computed using 2048 \textsc{Patchy} simulations. The left (right) plot gives the result for the `z1' (`z3') redshift slice, averaging together NGC and SGC measurements for visibility, as in Fig.\,\ref{fig: pk-bk-plot}. The correlation matrix is defined $R_{ij} \equiv C_{ij}/\sqrt{C_{ii}C_{jj}}$ for covariance $C_{ij}$. Here the first through third submatrices show the results from the power spectrum monopole, quadrupole and hexadecapole respectively, the fourth gives the result from the real-space power spectrum, $Q_0$, whilst the fifth gives that from the bispectrum. For clarity, we omit the BAO parameters, whose covariance with the windowed spectra can be found in \citep{2020JCAP...05..032P}. In each case, we include only $k$-modes used in the analysis below; \textit{i.e.}\ we use $0.01\leq k\leq 0.20$ for $P_\ell$, $0.20\leq k\leq 0.4$ for $Q_0$ and $0.01\leq k\leq 0.08$ for $B_0$ (in $\hMpc$ units). In general, there is little off-diagonal correlation, except for the $Q_0$ statistic which involves smaller scale information. Note that the correlation is less than conventionally found, since our measurements are not window convolved; however, the `z1' and `z3' covariances appear somewhat different, due to both the different redshift, and selection functions of the two. }
    \label{fig: cov-plot}
\end{figure}

Fig.\,\ref{fig: cov-plot} shows the corresponding correlation matrices, including all datasets considered in this work. Notably, we do not find strong correlations between the various statistics, aside from those in $P_\ell(k)$ and $P_{\ell'}(k)$, which are expected from linear theory. That the correlations are weaker than those found previously \citep{2020JCAP...05..032P,2017MNRAS.465.1757G,2017MNRAS.465.1757G} is also of no surprise: much of the large-scale correlations is imprinted by the window function, and will be substantially reduced by the window-free estimators of \S\ref{sec: estimators} \citep{2021PhRvD.103j3504P,2021arXiv210706287P}. We do however observe stronger correlations between bins in the real-space power spectrum $Q_0$: this occurs since this statistic includes smaller scales, whence the off-diagonal trispectrum covariance becomes important. Since we use only mock-based covariances, such higher-order couplings are naturally included in the analysis.

\subsection{Baseline Analysis}\label{subsec: baseline}

\subsubsection{Results}
\begin{figure*}[htb!]
\centering
\includegraphics[width=0.8\textwidth]{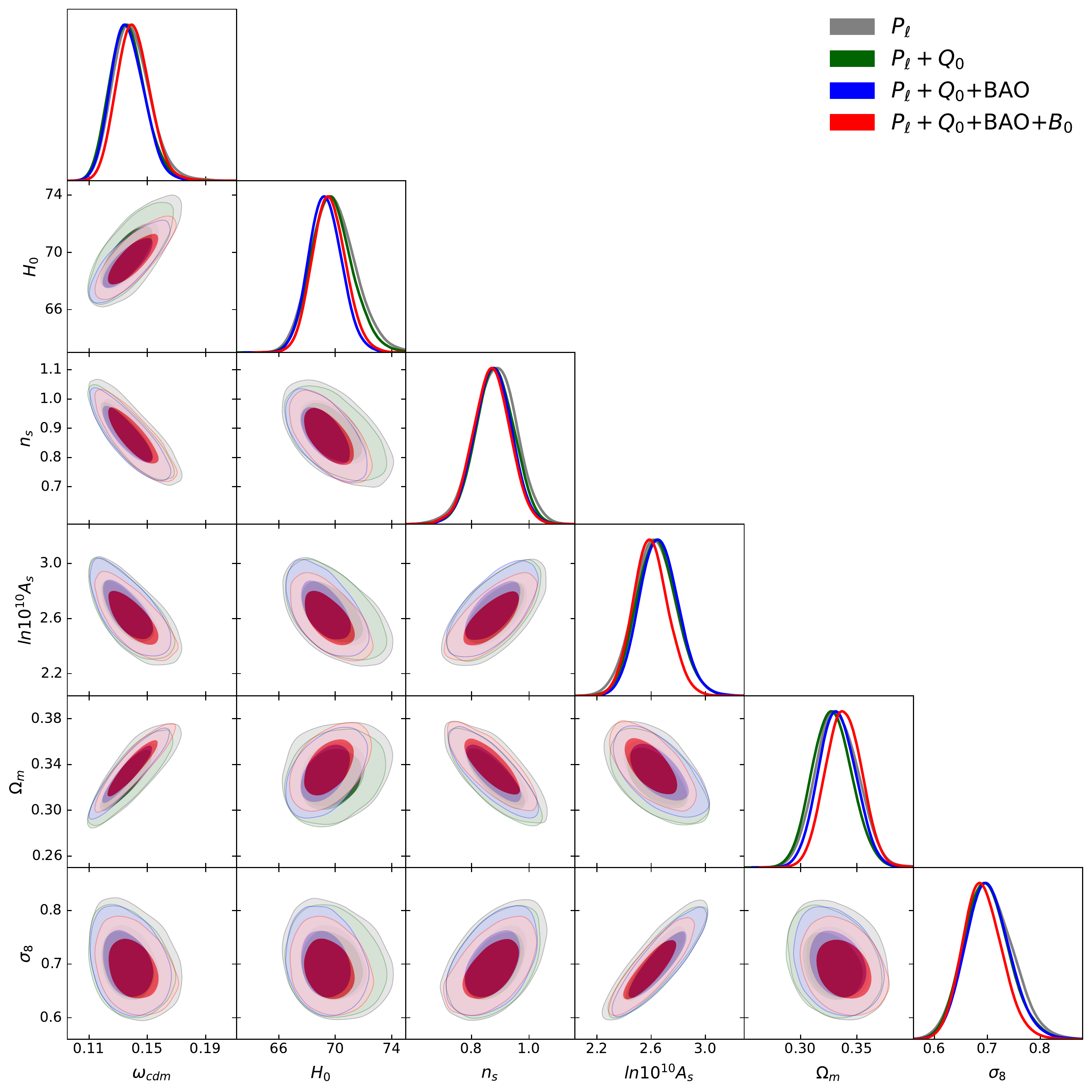}
\caption{Cosmological parameter constraints from the $\Lambda$CDM analysis of BOSS data, including the power spectrum ($P_\ell$), real-space power spectrum analog ($Q_0$), BAO parameters from reconstructed spectra (BAO), and the bispectrum monopole ($B_0$). The marginalized constraints are given in Tab.\,\ref{tab: free-ns}. As found previously, the addition of $Q_0$ gives a slight decrease in the posterior volume (limited mostly by shot-noise), whilst BAO parameters help to reduce the $H_0$ contour, and the bispectrum gives a $13\%$ reduction in the $\sigma_8$ error-bar. Notably, the spectral slope is poorly constrained and degenerate with other parameters; results with a \textit{Planck} prior on $n_s$ are shown in Fig.\,\ref{fig: boss-fix-ns}.}\label{fig: boss-free-ns}
\end{figure*}

We begin by analyzing the power spectrum, BAO parameters, and bispectrum in the manner described above, with all cosmological parameters (except the neutrino mass) free. The main results of this are displayed in Fig.\,\ref{fig: boss-free-ns} (and summarized in Fig.\,\ref{fig: main-results}), with the associated parameter confidence intervals given in Tab.\,\ref{tab: free-ns}, with additional results (including best-fits, 95\% confidence levels, and full corner plots including bias parameters) given in Appendix \ref{appen: full-constraints}. As more datasets are included in the analysis, we find that the mean parameter values are generally consistent, but the error-bars reduce. 

The addition of $Q_0$ gives a slight improvement in the precision with which a variety of parameters can be measured; this arises due to the inclusion of smaller-scale power spectrum information, the importance of which is limited by the high BOSS shot-noise \citep{2021arXiv211000006I}. When BAO parameters are included, the $H_0$ contour shrinks by approximately $20\%$, matching the conclusion of \citep{2020JCAP...05..032P}. Here, the extra constraining power is sourced by higher-order statistics, whose sound-horizon information is transferred to the power spectrum, following BAO reconstruction \citep{2015PhRvD..92l3522S}. Finally, when we add the bispectrum monopole, the $\sigma_8$ posterior shrinks by $13\%$ (in accordance with the $N$-body results of \citep{2021arXiv211010161I}), though other projections remain consistent. In this case, information arises from broadband features in the bispectrum, and allows for degeneracy breaking, which acts to tight the constraint on the fluctuation amplitude. Notably, the $H_0$ posterior is not tightened - this occurs since the large-scale bispectrum does not include BAO signatures (since we truncate at $k = 0.08\hMpc$). Information present on smaller scales is already included via the BAO parameters (see \citep{2018MNRAS.478.4500P} for a detection of this from the three-point function alone). Although the bispectrum is not found to give particularly large improvements in the error-bars of cosmological parameters, this is primarily due to our conservative choices of $k_{\rm max}$ and bias parameter priors, and would change if one accepted a larger systematic error budget or developed the theoretical modeling further \citep{2021arXiv211010161I}. If one is instead interested in galaxy formation physics, the bispectrum's addition improves constraints significantly; this is discussed in \S\ref{subsec: bias-results}.

\subsubsection{Cosmological Interpretation}\label{subsec: comparison}
The parameter contours found above can be straightforwardly compared to those of other analyses, both making use of LSS data, and other sources. For the Hubble constant, we find a marginalized constraint of $H_0 = 69.6^{+1.1}_{-1.3}\hun$ when all datasets are combined. This is consistent with that reported in \citep{2021arXiv211010161I} (using $P_\ell$ and $Q_0$ information), but somewhat higher than the results of \citep{2020JCAP...05..042I,2020JCAP...05..032P}. The difference is thought to arise from the power spectrum measurements themselves: in this work, and \citep{2021arXiv211010161I}, we do not use public data products, but instead remeasure the spectra from scratch (cf.\,\S\ref{subsec: pk-analysis-comparison}). Here, this is achieved via window-free estimators, which \S\ref{sec: consistency-tests} show to give robust parameter constraints. Our $H_0$ posterior is somewhat lower than those obtained from the Cepheid-calibrated distance ladder \citep{2021ApJ...908L...6R}: $H_0^{\rm SH0ES} = 73.2\pm 1.3\hun$, though, given the large error-bars, it is difficult to draw any firm conclusions. The $\Omega_m$ posterior is consistent with that of the Pantheon sample: $\Omega_m^{\rm Pantheon} = 0.298\pm 0.022$ \citep{2018ApJ...859..101S}, and the $n_s$ posterior is broad, but in agreement with that from \textit{Planck}: $n_s^{\rm\textit{Planck}} = 0.9649\pm0.0042$.

The $\sigma_8$ results are of particular interest. Our full analysis finds $\sigma_8 = 0.69\pm0.04$ (or $\sigma_8=0.70\pm 0.05$, using the power spectrum multipoles alone), which is somewhat different from that of previous full-shape power spectrum analyses \citep{2020JCAP...05..042I,2021PhRvD.103b3538P,2020JCAP...05..005D,2021arXiv211006969K}. Comparison with the old analyses in not straightforward, since they differ in many aspects, such as the inclusion of the hexadecapole moment in the data and window function model, the choice of sampling parameters, e.g. whether to use $A_s/A_{s,\rm Planck}$ or $\ln (10^{10}A_s)$, and finally, the priors on nuisance parameters. If we analyze the power spectrum data used by the former works with the same theory model and data cuts as our baseline analysis (see \S\ref{subsec: pk-analysis-comparison}), and consistently treat the window function, we find $\sigma_8=0.66\pm 0.05$, which is significantly lower than the \textit{Planck} value, and $1\sigma$ below our current bound.

This difference is entirely caused by an error in the public BOSS power spectra,\footnote{Available at \href{https://fbeutler.github.io/hub/boss_papers.html}{fbeutler.GitHub.io/hub/boss\_papers.html}} whereupon the spectra were incorrectly suppressed by a constant factor $\approx 10\%$. This occurred due to an invalid approximation in the power spectrum normalization: the geometric factor $A \equiv \int d\vr\,\bar{n}^2(\vr)w_{\rm fkp}^2(\vr)$ appearing in the standard FKP power spectrum estimator was replaced by a sum over random particles:
\beq
    A_{\rm approx} = \frac{N_{\rm g}}{N_{\rm r}}\sum_{i\,\in\,\mathrm{randoms}} \bar n_g(z_i)w^2_{\rm fkp,i}\,,
\eeq
for approximate redshift dependence $\bar n_g(z_i)$, using a (weighted) total of $N_{\rm gal}$ galaxies and $N_{\rm rand}$ randoms \citep{2017MNRAS.464.3409B,2017MNRAS.466.2242B,2017MNRAS.465.1757G,2021arXiv211006969K}. This approximation is valid only at the $\approx 10\%$ level for the BOSS sample, and the power spectra should strictly be normalized by the $r\to0$ limit of the window function $W(r)$ (which is equal to the two-point correlation function of the random particles) \citep{2021JCAP...11..031B,2021MNRAS.501.5616D}.\footnote{We caution that this normalization is still used in the \textsc{nbodykit} code \citep{2018AJ....156..160H}, and was also present in the reanalysis of \citep{2021arXiv211000006I}. To correct for it, one must similarly normalize the window function by $A_{\rm approx}$, instead of the usual approach, which sets $\lim_{r\to0}W(r) = 1$.}  At linear order, $P(k,\mu) \propto \left[b+f\mu^2\right]^2\sigma^2_8$, thus this has the effect of reducing the value of $\sigma_8$ by $\sim 5\%$ relative to its true value. This resulted in a a series of works that found $\sigma_8$ to be in (erroneous) mild tension with \textit{Planck}. In our approach, we do not require the 2PCF of the mask to be computed explicitly, thus are not affected by such problems. Our results for $\sigma_8$ are consistent with those found in \citep{2021arXiv211005530C}, which uses the correctly normalized spectra.

Recent works have found a growing tendency for weak-lensing analyses to predict lower clustering amplitudes than that of \textit{Planck}; an anomaly that shows greater consistency between observables than the much-ballyhooed `$H_0$ tension'. In terms of the $S_8 \equiv \sigma_8\left(\Omega_m/0.3\right)^{0.5}$ parameter, \textit{Planck} finds $S_8^{Planck} = 0.832\pm0.012$ \citep{2020A&A...641A...6P}, whilst the latest weak lensing results from the latest Dark Energy Survey $3\times2$ analysis finds $S_8^{\rm DES} = 0.776\pm0.017$ \citep{2021arXiv210513549D}. An additional constraint is obtained from cross-correlating galaxy surveys with the CMB: this obtained $S_8 = 0.776\pm0.017$ for the unWISE sample \citep{2021arXiv210503421K}, and $S_8 = 0.73\pm0.03$ when using the DESI imaging survey \citep{2021arXiv211109898W}. Here, we find $S_8 = 0.734^{+0.035}_{-0.041}$; this is consistent with the other low-redshift observations, but differs from \textit{Planck} at the $2.5\sigma$ level. The upcoming tranche of spectroscopic and photometric survey data will be able to shed light on whether this is indeed a \textit{bona fide} discrepancy (the prospect of which is discussed in \S\ref{subsec: s8-physics}), or simply a statistical fluctuation.

\begin{table}[!t]
    \centering
    \rowcolors{2}{white}{vlightgray}
  \begin{tabular}{|c|cccc|cc|} \hline
    \textbf{Dataset} & $\omega_{\rm cdm}$ & $h$ & $\ln(10^{10}A_s)$ & $n_s$ & $\Omega_m$ & $\sigma_8$\\\hline
   $P_\ell(k)$ &  $\quad 0.139_{-0.015}^{+0.011}\quad $ & $\quad 0.699_{-0.017}^{+0.015}\quad $ & $\quad 2.63_{-0.16}^{+0.15}\quad $& $\quad 0.883_{-0.072}^{+0.076}\quad $ & $\quad 0.333_{-0.020}^{+0.019}\quad $ & $\quad 0.704_{-0.049}^{+0.044}\quad $\\
   $P_\ell(k)+Q_0(k)$ & $0.137_{-0.014}^{+0.011}$ & $0.698_{-0.016}^{+0.013}$ & $2.64_{-0.16}^{+0.14}$ & $0.880_{-0.068}^{+0.068}$ & $0.328_{-0.019}^{+0.017}$ & $0.699_{-0.046}^{+0.040}$\\
   $P_\ell(k)+Q_0(k)+\mathrm{BAO}$ & $0.137_{-0.013}^{+0.011}$ & $0.693_{-0.013}^{+0.011}$ & $2.66_{-0.15}^{+0.14}$ & $0.874_{-0.064}^{+0.067}$ & $0.333_{-0.018}^{+0.016}$ & $0.701_{-0.045}^{+0.040}$\\
  \quad$P_\ell(k)+Q_0(k)+\mathrm{BAO}+B_0$\quad\quad  & $0.141_{-0.013}^{+0.011}$ & $0.696_{-0.013}^{+0.011}$ & $2.60_{-0.14}^{+0.13}$ & $0.870_{-0.064}^{+0.067}$ & $0.338_{-0.017}^{+0.016}$ & $0.692_{-0.041}^{+0.035}$\\\hline
    \end{tabular}
    \caption{Cosmological parameter constraints from the $\Lambda$CDM analysis of BOSS using the power spectrum multipoles, real-space power spectrum, BAO parameters, and bispectrum. For each analysis, we display mean values and 68\% confidence intervals. A BBN prior on the physical baryon density $\omega_b$ is assumed in all cases, and the corresponding posterior is not displayed. The left group of parameters are those directly sampled in the MCMC chains, whilst those on the right are derived parameters. A corner plot of these results is given in Fig.\,\ref{fig: boss-free-ns}, and bias parameter constraints from the final two datasets are shown in Appendix \ref{appen: full-constraints}.}
\label{tab: free-ns}
\end{table}

\subsection{Analysis with Fixed $n_s$}
\subsubsection{Results}
Spectroscopic surveys generally lead to poor constraints on the primordial slope, $n_s$, due to the limited survey size, and thus the resolution of large-scale modes. For this reason, many analyses have opted to fix $n_s$ by imposing a tight Gaussian prior on $n_s$, with width given by the \textit{Planck} constraints \citep{2020A&A...641A...6P}. To allow straightforward comparison with previous works, we additionally perform the analysis described above using the $n_s$-prior, the results of which are shown in Fig.\,\ref{fig: boss-fix-ns} and Tab.\,\ref{tab:fix-ns}, with additional constraints displayed in Appendix \ref{appen: full-constraints}.

Our conclusions in this case are similar to before: the inclusion of $Q_0$ leads to a slight reduction in parameter variances, BAO tightens the $H_0$ posterior by $\approx 20\%$, and $B_0$ sharpens the $\sigma_8$ constraint by $13\%$. Including a highly restrictive prior on $n_s$ reduces a number of parameter degeneracies, improving the precision with which $\Omega_m$ can be measured by almost $40\%$. For $H_0$, we find a posterior of $68.3^{+0.8}_{-0.9}\hun$ when all datasets are included, which is closer to the \textit{Planck} result, and further from that of Cepheid-calibrated supernovae. For $\sigma_8$, the marginalized constraint becomes $0.72^{+0.03}_{-0.04}$, now with an error below $<5\%$, whilst $S_8$ is given by $0.751\pm0.039$, now within $2\sigma$ of \textit{Planck}. We caution that these shifts are a natural phenomenon of non-Gaussian posteriors, and are not an indication of some failure of the cosmological model. Indeed, from Fig.\,\ref{fig: boss-free-ns} it is evident that the value of $n_s$ preferred by \textit{Planck} is higher than that directly inferred from the BOSS data. As such, imposing the \textit{Planck} $n_s$ prior pulls $\sigma_8$ to larger values due to the variables' positive correlation. This explains the small upward shift in $\sigma_8$ and $S_8$, which reduces the tension with \textit{Planck}.

\subsubsection{Comparison to Other Power Spectrum Analyses}\label{subsec: pk-analysis-comparison}
As an additional test of our pipeline, we compare our results to those obtained using the publicly available BOSS DR12 power spectrum measurements (corrected for the normalization issue discussed in \S\ref{subsec: comparison}). For this purpose, we reanalyze the public data using the same theory model and data-cuts described above, but restricting to the power spectrum multipoles, and including the window function in the theory model (as in \citep{2020JCAP...05..042I}). This ensures that the analyses differ only due to the underlying power spectra and covariance matrix. 

\begin{figure*}[htb!]
\centering
\includegraphics[width=0.6\textwidth]{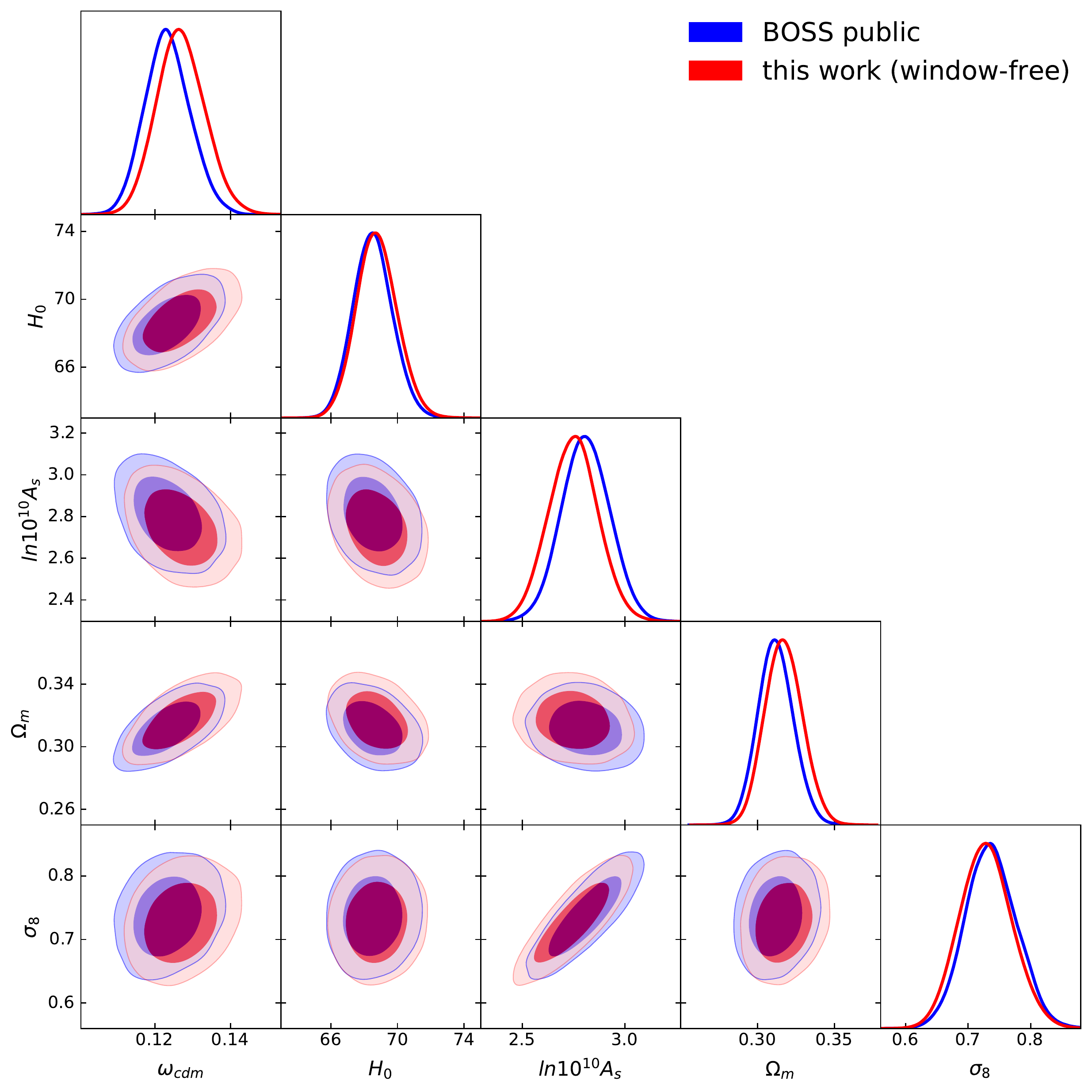}
\caption{Comparison of cosmological posteriors obtained from an analysis of the BOSS power spectrum using the window-free estimators of this work to that using public (windowed) BOSS data. We adopt the same theory model and scale-cuts in each case, such that differences can arise only due to the underlying data and associated covariance matrices. Our results are generally in good agreement, though our analyses find a slightly increased $\omega_{\rm cdm}$ (and thus a reduced and $A_s$, due to a strong anticorrelation). This is likely caused by the different window function treatments and $k$-space binning.}\label{fig:compar}
\end{figure*}

As shown in Tab.\,\ref{tab:fix-ns} and Fig.\,\ref{fig:compar}, we find a close-to-perfect agreement between the two analyses, with the only noticeable difference being a $\lesssim 0.5\sigma$ downward shift of $\omega_{cdm}$, which, in turn, pulls $A_s$ upwards due to an anticorrelation.\footnote{Note that $\omega_{cdm}$ is measured directly from the broadband shape of the monopole, whilst $A_s$ is primarily derived from $f\sigma_8$, rather than being measured directly from the full-shape data \citep{2020JCAP...05..042I}.} The shift is insignificant at the level of the current data, and could be caused either by a inaccuracy in the window function modeling in the public data, or a residual measurement systematic in the new approach (though the latter was found to produce highly accurate $\omega_{\rm cdm}$ estimates in \S\ref{sec: consistency-tests}). This shift is also similar to the $0.5\sigma$ bias on $\omega_{\rm cdm}$ found to arise from noise in the sample covariance matrix in \citep{2021PhRvD.103d3508P,Wadekar:2020hax}. In this regard, it is important to note that the public covariance matrices use $k$-bins of width $\Delta k = 0.01\hMpc$, which is twice as large as our choice. Although we do not expect the binning to affect the power spectrum model, it will modify the noise properties of the covariance, since it changes the number of data points for a given number of mocks. A thorough investigation of this issue goes beyond the scope of this paper. 

We may also compare the results of this section with other recent full-shape analyses that include the \textit{Planck} prior on the spectral tilt. In particular, our results are in good agreement with \citep{2021arXiv211005530C}, which analyzed the (corrected) public BOSS DR12 power spectra using a similar perturbation theory model and data cuts. \resub{In particular, the former work found $\sigma_8=0.733\pm 0.047$, in close agreement with Tab.\,\ref{tab: all-constraints-fix-ns}. Given that the two methods use different window function treatments and distinct flavors of perturbation theory, this is a useful validation of our power spectrum analysis.}

An additional comparison is between our work and that of \cite{2021arXiv211006969K}, which used an emulator-based approach to analyze the public BOSS data. Although the underlying model is very different to our approach, the posterior means and variances agree quite well except for $\sigma_8$, for which \cite{2021arXiv211006969K} finds a somewhat larger value, \resub{inconsistent with \cite{2021arXiv211005530C} and that of this work}. \resub{As pointed out by \cite{2021arXiv211005530C} (and \S\ref{subsec: s8-physics}), the $\sigma_8$ posterior is determined by large scales, $k\lesssim0.1\hMpc$, thus the difference between our results is unlikely to be due to the treatment of non-linear corrections.} Given the significant difference between our methodologies, a thorough investigation of this discrepancy requires significant work. We hope to address these shifts further in future.

\begin{figure*}[htb!]
\centering
\includegraphics[width=0.7\textwidth]{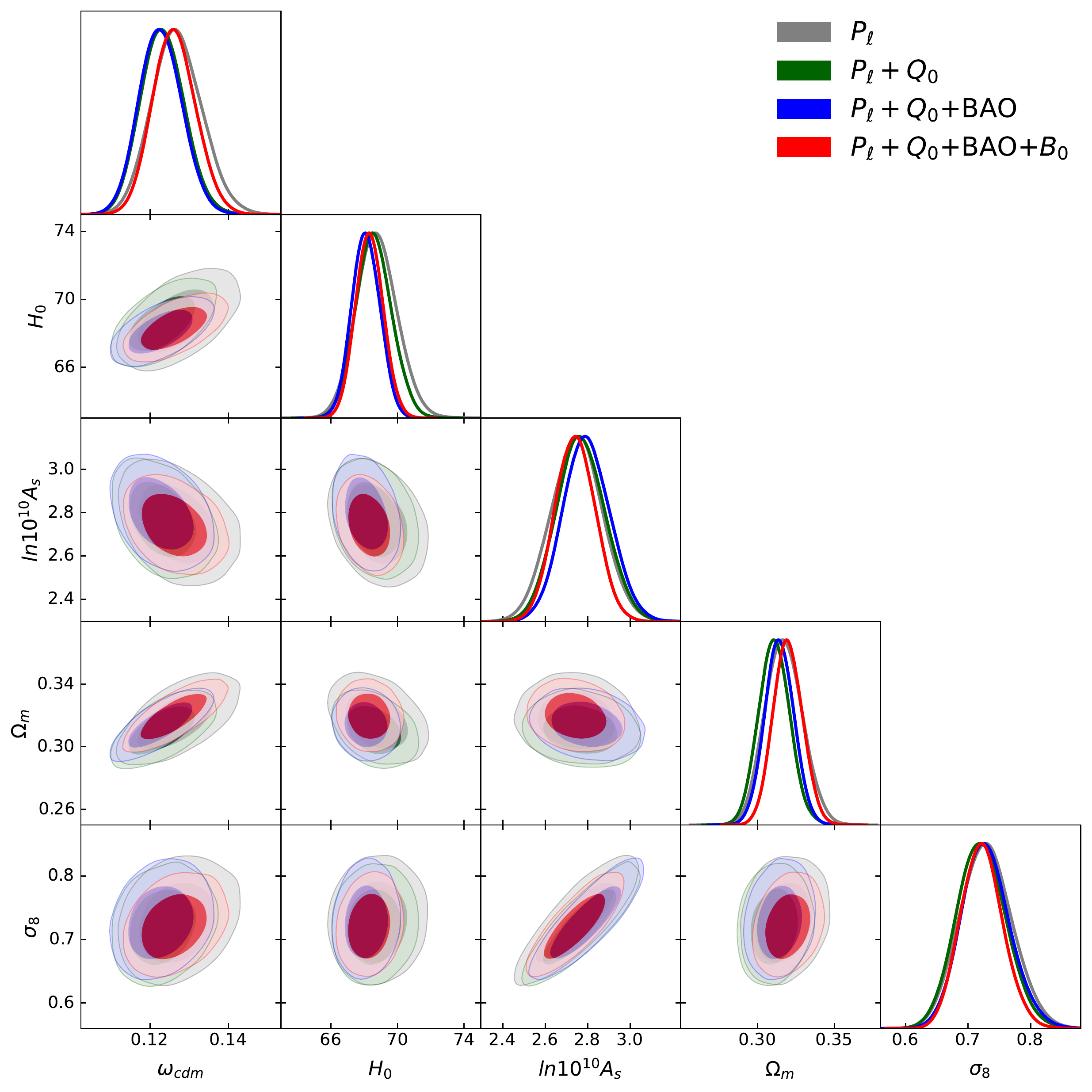}
\caption{As Fig.\,\ref{fig: boss-free-ns}, but with a \textit{Planck} prior on the spectra slope $n_s$. Our conclusions are broadly consistent with that of the free-$n_s$ analysis, but feature somewhat tighter parameter constraints. The corresponding marginalized posteriors are shown in Tab.\,\ref{tab:fix-ns}.}
\label{fig: boss-fix-ns}
\end{figure*}


\begin{table}[!t]
    \centering
    \rowcolors{2}{white}{vlightgray}
  \begin{tabular}{|c|ccc|cc|} \hline
    \textbf{Dataset} & $\omega_{\rm cdm}$ & $h$ & $\ln(10^{10}A_s)$ & $\Omega_m$ & $\sigma_8$\\\hline
   $P_\ell(k)$, public & $\quad 0.1233_{-0.0065}^{+0.0058}\quad $ & $\quad 0.685_{-0.013}^{+0.011}\quad $& $\quad 2.81_{-0.12}^{+0.12}\quad $& $\quad 0.312_{-0.012}^{+0.011}\quad $& $\quad 0.737_{-0.044}^{+0.040}\quad $ \\
   $P_\ell(k)$ & $0.1268_{-0.0068}^{+0.0062}$ & $0.688_{-0.013}^{+0.012}$& $2.75_{-0.13}^{+0.12}$& $0.317_{-0.013}^{+0.012}$& $0.729_{-0.045}^{+0.040}$ \\
   $P_\ell(k)+Q_0(k)$ & $0.1232_{-0.0058}^{+0.0054}$ & $0.686_{-0.011}^{+0.011}$& $2.77_{-0.12}^{+0.11}$& $0.311_{-0.010}^{+0.010}$&$0.722_{-0.042}^{+0.037}$\\
   $P_\ell(k)+Q_0(k)+\mathrm{BAO}$ & $0.1227_{-0.0059}^{+0.0053}$ & $0.6811_{-0.0089}^{+0.0083}$& $2.80_{-0.12}^{+0.11}$& $0.314_{-0.010}^{+0.010}$&  $0.729_{-0.042}^{+0.036}$\\
  \quad$P_\ell(k)+Q_0(k)+\mathrm{BAO}+B_0$\quad\quad & $0.1262_{-0.0059}^{+0.0053}$ & $0.6831_{-0.0086}^{+0.0083}$& $2.741_{-0.098}^{+0.096}$& $0.320_{-0.010}^{+0.010}$& $0.722_{-0.036}^{+0.032}$\\\hline
    \end{tabular}
    \caption{As Tab.\,\ref{tab: free-ns}, but including a \textit{Planck} prior on $n_s$. The corresponding corner plot is given in Fig.\,\ref{fig: boss-fix-ns}. We additionally include the results of a windowed analysis of the public BOSS power spectrum (top line); this is described in \S\ref{subsec: pk-analysis-comparison} and provides a useful test of our analysis pipeline.}
\label{tab:fix-ns}
\end{table}

\subsection{On The Plausibility of New Physics}\label{subsec: s8-physics}

\begin{figure}
    \centering
    \includegraphics[width=0.49\textwidth]{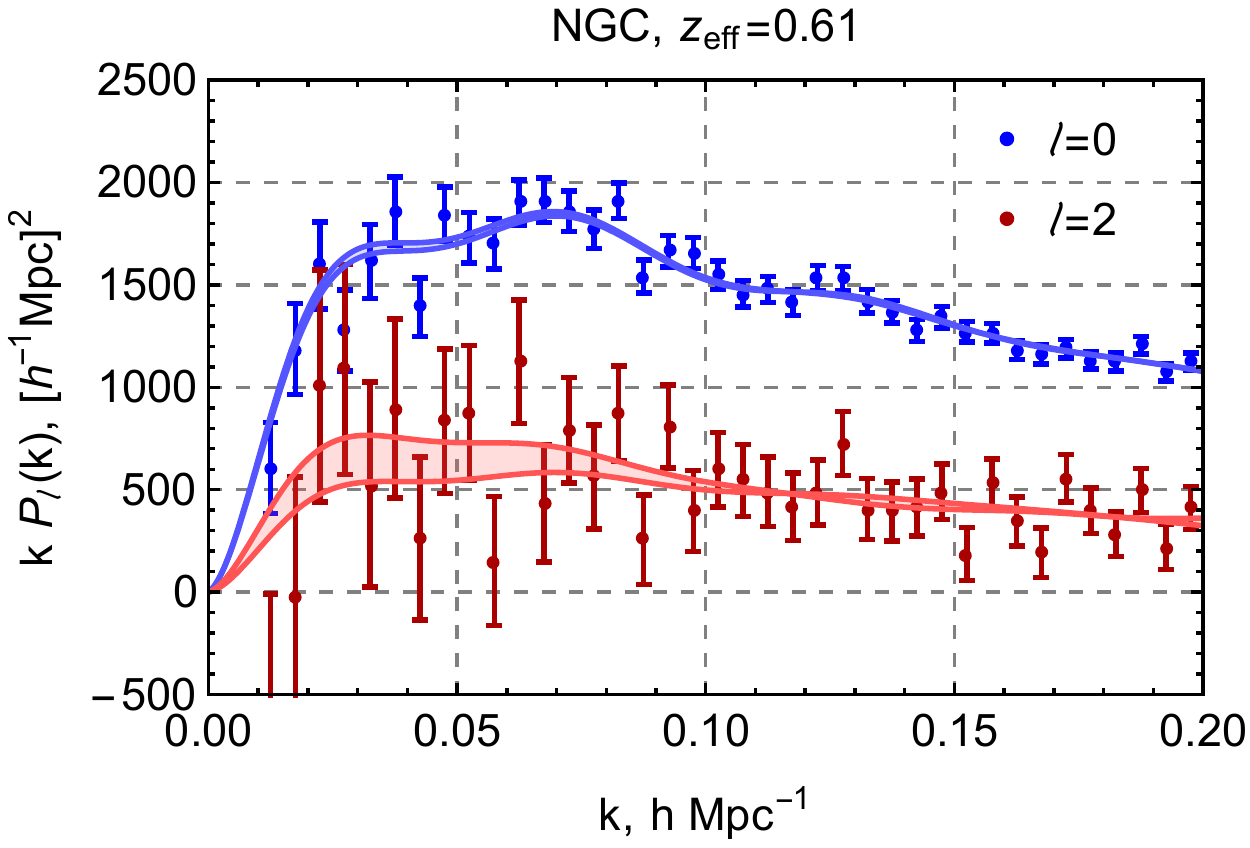}
        \includegraphics[width=0.49\textwidth]{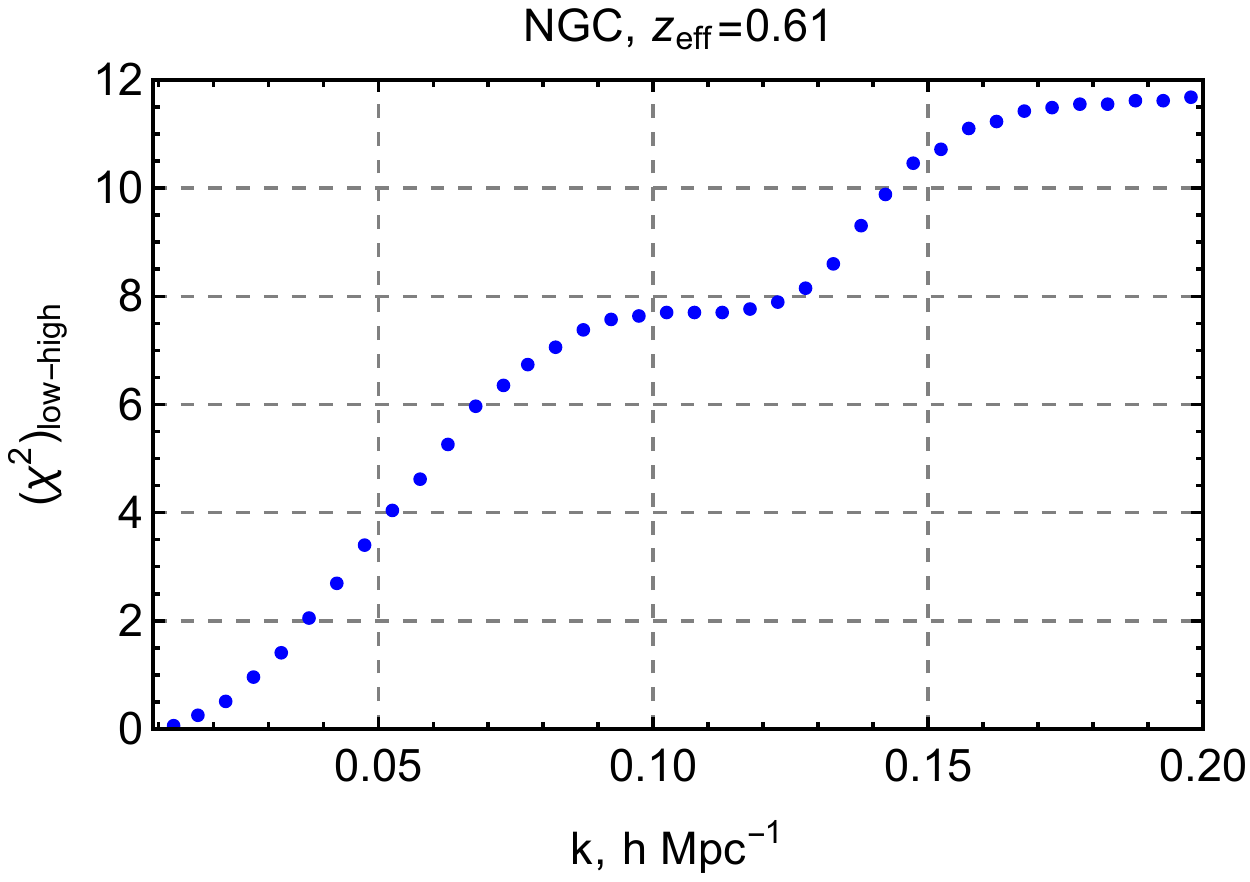}
    \caption{\textit{Left panel:} Variations in the power spectrum multipoles induced by changing $\sigma_8$ to values with $\Delta\chi^2 = 1$ compared to the best-fit model, corresponding to $\sigma_8=0.57$ (bottom) and $\sigma_8=0.83$ (top), keeping other parameters fixed. For clarity, we plot only the power spectrum monopole (blue) and quadrupole (red) from the NGC `z3' region, with error-bars obtained from the \textsc{Patchy} mocks. Note that there are non-trivial correlations between data-points. \textit{Right panel:} Cumulative $\chi^2$ difference between the theory models with high and low $\sigma_8$ as a function of scale, \textit{i.e.} 
    $\left[P_{\sigma_8 = 0.57}(k_i)-P_{\sigma_8 = 0.83}(k_i)\right]\mathsf{C}_{ij}^{-1}\left[P_{\sigma_8 = 0.57}(k_j)-P_{\sigma_8 = 0.83}(k_j)\right]$.
    Notably, this is dominated by scales with $k\lesssim0.1\hMpc$. 
    }
    \label{fig:s8}
\end{figure}

To shed light on possible tensions between our results and those of \textit{Planck}, it is instructive to consider how our best-fit model depends on $\sigma_8$. This is shown in Fig.\,\ref{fig:s8}: in the left panel we plot the power spectrum monopole and quadrupole for two values of $\sigma_8$, chosen such that the resulting model has a $\Delta\chi^2$ of unity with respect to the best-fit. Notably, we observe significant changes only in the large-scale quadrupole ($k\lesssim 0.1\hMpc$), implying that it is this region that drives our $\sigma_8$ constraints. In the right panel of Fig.\,\ref{fig:s8} we show the cumulative $\chi^2$ between the models with $\sigma_8=0.57$ and $\sigma_8=0.83$ as a function of $k_{\rm max}$. This metric is equivalent to the signal-to-noise ratio corresponding to a detection of the difference between the two models. The total $\chi^2$ deviation between the two models is $\approx 12$, the majority of which is accumulated from $k<0.1~\hMpc$: the $\chi^2$ difference grows monotonically in this range, reaffirming our notion that these scales are most important for distinguishing between models with different clustering amplitudes. Furthermore, it can be shown that, to obtain $\sigma_8 = 0.83$ within our $P_\ell$-only analysis, unconventional values of the bias parameters are required, such as $b_2 \approx -2$, which are generally ruled out when the bispectrum is also included. In contrast, the bias parameters associated with our best-fit $\sigma_8$ value are generally consistent with the expected halo bias relations discussed in \S\ref{subsec: bias-results}.

A wide variety of new physics models have been proposed to alleviate the so-called $S_8$-tension, including via decaying dark matter, modified gravity, interacting dark energy and sterile neutrinos \citep[e.g.,][]{2016A&A...594A..14P,2020arXiv200809615A,2020JCAP...12..016H,2014PhRvL.112e1303B,2020PDU....3000666D} (see \cite{DiValentino:2020vvd} for a recent review). For most of these models, one would expect a suppression of power at large $k$, beyond the characteristic scale of the phenomenon. A typical example is the neutrino free-streaming scale $k_{\rm fs}\approx 0.1~\hMpc$ for $m_\nu\approx0.1~$eV~\citep{Lesgourgues:2006nd}. Obtaining a model capable of changing only the large-scale quadrupole is a more difficult feat, and may require exotic physics. For instance, the presence of this large-scale discrepancy prohibits the $\sigma_8$ difference between BOSS galaxies and the \textit{Planck} CMB being fit by a massive neutrino \cite{Ivanov:2019hqk}.

An alternative explanation is that there are systematics in our modeling. Given the results of Fig.\,\ref{fig:s8}, it seems unlikely that omitted higher-order effects could generate such a shift: $\sigma_8$ is primarily set by large scales, whose character is well described by linear physics. One possibility is that selection effects could influence the $\sigma_8$ posterior. These arise from anisotropic assembly biases, which violate the symmetry arguments used to construct the bias expansion, with the effect of modifying the large-scale quadrupole \citep{2009MNRAS.399.1074H}. Estimates of the magnitude of this effect range from close to zero \citep{Singh:2020cvu} to highly significant \citep{2020JCAP...10..058O}, suggesting the need for further study. Given that our $S_8$ results are consistent with those from weak lensing probes (which do not have access to redshift-space information), it seems likely that such selection effects do not significantly affect the $\sigma_8$ constraints found herein BOSS data. The combination of the above results promote the conclusion that this discrepancy arises simply due to noise fluctuations.

\subsection{Testing Bias Relations}\label{subsec: bias-results}
Galaxy bias parameters are a key part of any perturbative model, yet they enter the power spectrum in a degenerate manner that makes their individual determination difficult (with the exception of linear bias). As a result, previous constraints on the quadratic and tidal biases, $b_2$, and $b_{\mathcal{G}_2}$ were prior-dominated \citep[e.g.,][]{2020JCAP...05..042I}. In contrast, $b_2$ and $b_{\mathcal{G}_2}$ appear in the tree-level bispectrum model accompanied by different shapes, allowing them to be directly constrained from data \citep{2021arXiv211010161I}. For dark-matter halos, there are well-known relations between bias parameters \citep{2012PhRvD..86h3540B,2018PhR...733....1D,2020PhRvD.102j3530E,2016JCAP...02..018L,2018JCAP...09..008L,2018JCAP...07..029A}; the inclusion of bispectrum information allows one to study whether these hold also for observed galaxies.

\begin{table}
    \centering
   \rowcolors{2}{white}{vlightgray}
  \begin{tabular}{|c|cccc|cccc|}
  \hline
  & \multicolumn{4}{c|}{$\bm{P_\ell + Q_0 + \mathrm{BAO}}$} & \multicolumn{4}{c|}{$\bm{P_\ell + Q_0 + \mathrm{BAO} + B_0}$}\\\hline
  \quad \textbf{Parameter}\quad\quad & best-fit & mean$\,\pm\,\sigma$ & \quad 95\% lower \quad & \quad 95\% upper \quad & best-fit & mean$\,\pm\,\sigma$ & \quad 95\% lower \quad & \quad 95\% upper \quad \\ \hline
$b^{(1)}_{1 }$ &$2.12$ & $2.21_{-0.16}^{+0.15}$ & $1.90$ & $2.52$ &$2.26$ & $2.45_{-0.14}^{+0.15}$ & $2.15$ & $2.73$ \\
$b^{(1)}_{2 }$ &$-2.64$ & $-2.32_{-1.20}^{+0.74}$ & $-4.37$ & $-0.134$ &$0.273$ & $1.10_{-1.40}^{+1.00}$ & $-1.17$ & $3.66$  \\
$b^{(1)}_{{\mathcal G_2} }$ &$-0.106$ & $-0.12_{-0.48}^{+0.44}$ & $-1.03$ & $0.808$ &$-0.362$ & $-0.16_{-0.36}^{+0.27}$ & $-0.806$ & $0.545$ \\
$b^{(2)}_{1 }$ &$2.34$ & $2.40_{-0.16}^{+0.16}$ & $2.09$ & $2.71$ &$2.41$ & $2.58_{-0.15}^{+0.15}$ & $2.28$ & $2.88$ \\
$b^{(2)}_{2 }$ &$-0.834$ & $0.42_{-2.6}^{+1.6}$ & $-3.13$ & $4.55$ &$0.341$ & $0.93_{-1.4}^{+1.1}$ & $-1.49$ & $3.55$ \\
$b^{(2)}_{{\mathcal G_2} }$ &$-0.711$ & \quad$-0.135_{-0.57}^{+0.53}$ & $-1.20$ & $0.976$
&$-0.362$ & $-0.161_{-0.36}^{+0.27}$ & $-0.806$ & $0.545$ \\
$b^{(3)}_{1 }$ &$2.04$ & $2.11_{-0.15}^{+0.14}$ & $1.83$ & $2.40$ &$2.13$ & $2.30_{-0.13}^{+0.13}$ & $2.03$ & $2.57$ \\
$b^{(3)}_{2 }$ &$-1.12$ & $-1.07_{-1.50}^{+0.71}$ & $-3.33$ & $1.83$ &$-0.024$ & $0.375_{-0.82}^{+0.62}$ & $-1.03$ & $1.88$ \\
$b^{(3)}_{{\mathcal G_2} }$ &$-0.386$ & $-0.409_{-0.42}^{+0.4}$ & $-1.24$ & $0.443$ &$-0.403$ & $-0.394_{-0.20}^{+0.18}$ & $-0.788$ & $0.0072$ \\
$b^{(4)}_{1 }$ &$2.10$ & $2.15_{-0.15}^{+0.14}$ & $1.86$ & $2.44$ &$2.16$ & $2.34_{-0.14}^{+0.14}$ & $2.06$ & $2.61$ \\
$b^{(4)}_{2 }$ &$-1.86$ & $-0.689_{-2.00}^{+0.79}$ & $-3.33$ & $3.09$ &$0.124$ & $0.478_{-0.92}^{+0.70}$ & $-1.10$ & $2.21$ \\
$b^{(4)}_{{\mathcal G_2} }$ &\quad$-0.238$ & $0.0305_{-0.47}^{+0.44}$ & $-0.879$ & $0.952$ &$-0.203$ & $-0.014_{-0.29}^{+0.25}$ & $-0.556$ & $0.550$ \\\hline
 \end{tabular}
 \caption{Posterior values for the linear and quadratic bias parameters extracted from a power spectrum (left) or power spectrum plus bispectrum (right) analysis without restrictive priors on $b_2$ and $b_{\mathcal{G}_2}$. The superscripts on bias parameters indicate the sample, in the order NGC z3, SGC z3, NGC z1, SGC z1. The corresponding two-dimensional posterior is shown in Fig.\,\ref{fig:bias2d}.}\label{tab:bias}\end{table}

Marginalized constraints on the galaxy bias parameters for the four BOSS data chunks are given in Tab.\,\ref{tab: all-constraints-free-ns}\,\&\,\ref{tab: all-constraints-fix-ns} from the power spectrum and bispectrum, optionally including the \textit{Planck} prior on $n_s$. The inclusion of the bispectrum sharpens second-order bias parameters by up to $30\%$, with a much greater improvement seen than for the cosmological parameters. In the particular degeneracy direction $b_{\mathcal{G}_2}-b_2/2.3$, the bispectrum narrows the posterior by a factor of two. In reality, this comparison underestimates the utility of the bispectrum since, in its absence, the bias parameter constraints are dominated by restrictive physical priors. If these are not imposed, the posterior widths of the second-order biases are much larger \citep[cf.\,][]{2020JCAP...05..042I}. This implies that the actual improvement from the bispectrum can be significantly larger, as found in \citep{2021arXiv211010161I}. 

\begin{figure}
    \centering
    \includegraphics[width=\textwidth]{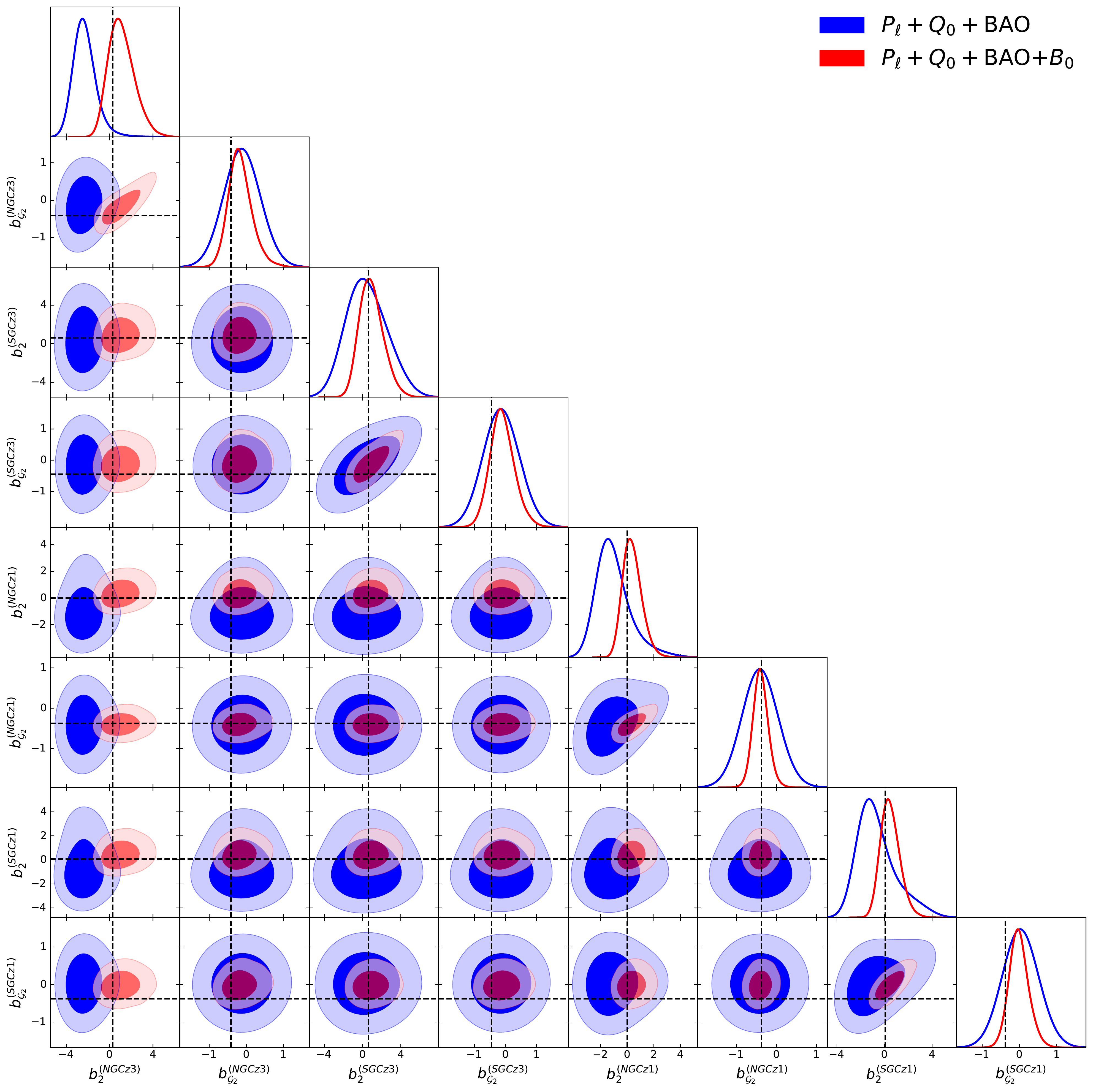}
    \caption{Quadratic bias parameters from an analysis of the BOSS DR12 power spectrum (blue, including the real-space proxy and BAO parameters) and power spectrum plus bispectrum (right), without imposing restrictive priors. Vertical and horizontal lines mark the predictions of the dark matter halo bias relations for the optimal values of $b_1$ for each sample, using the formalisms described in the main text. The associated marginalized constraints are shown in Tab.\,\ref{tab:bias}. Notably, the large-scale bispectrum significantly sharpens constraints on the second-order biases. 
    }
    \label{fig:bias2d}
\end{figure}

To explore this, we have rerun our fixed-$n_s$ analysis without restrictive priors on $b_2$ and $b_{\mathcal{G}_2}$, changing them to flat priors with infinite support. The corresponding constraints are shown in Fig.~\ref{fig:bias2d} and in Table.~\ref{tab:bias}. Dashed lines in Fig.~\ref{fig:bias2d} indicate the predictions based on the dark matter halo bias relations, as above. In the absence of the bispectrum dataset, we find highly non-Gaussian posteriors, with $b_2$ constraints shifted towards large negative values for almost all data chunks. This behaviour suggests that the posteriors are not dominated by the data. Indeed, without the bispectrum there are many unbroken degeneracies between the biases and other nuisance parameters, making $b_2$ and $b_{\mathcal{G}_2}$ quite sensitive to the choice of priors. When the bispectrum is included, the posterior widths shrink dramatically in all cases, and the distribution becomes closer to Gaussian. For the NGCz3 chunk we find that the power spectrum results are somewhat biased with respect to the dark matter coevolution prediction; this disappears when the bispectrum is included, indicating that it is a statistical fluctuation.

\begin{figure}
	\centering
	\includegraphics[width=0.49\textwidth]{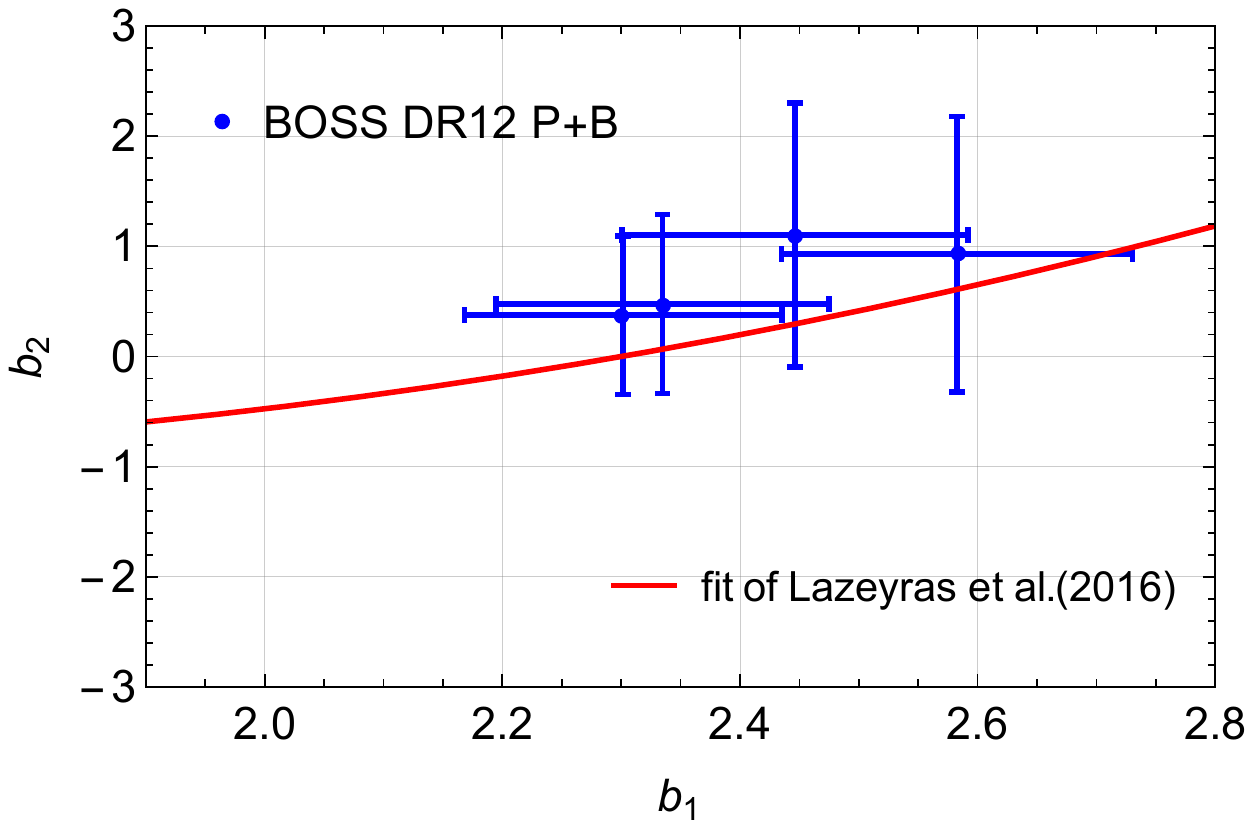}
	\includegraphics[width=0.49\textwidth]{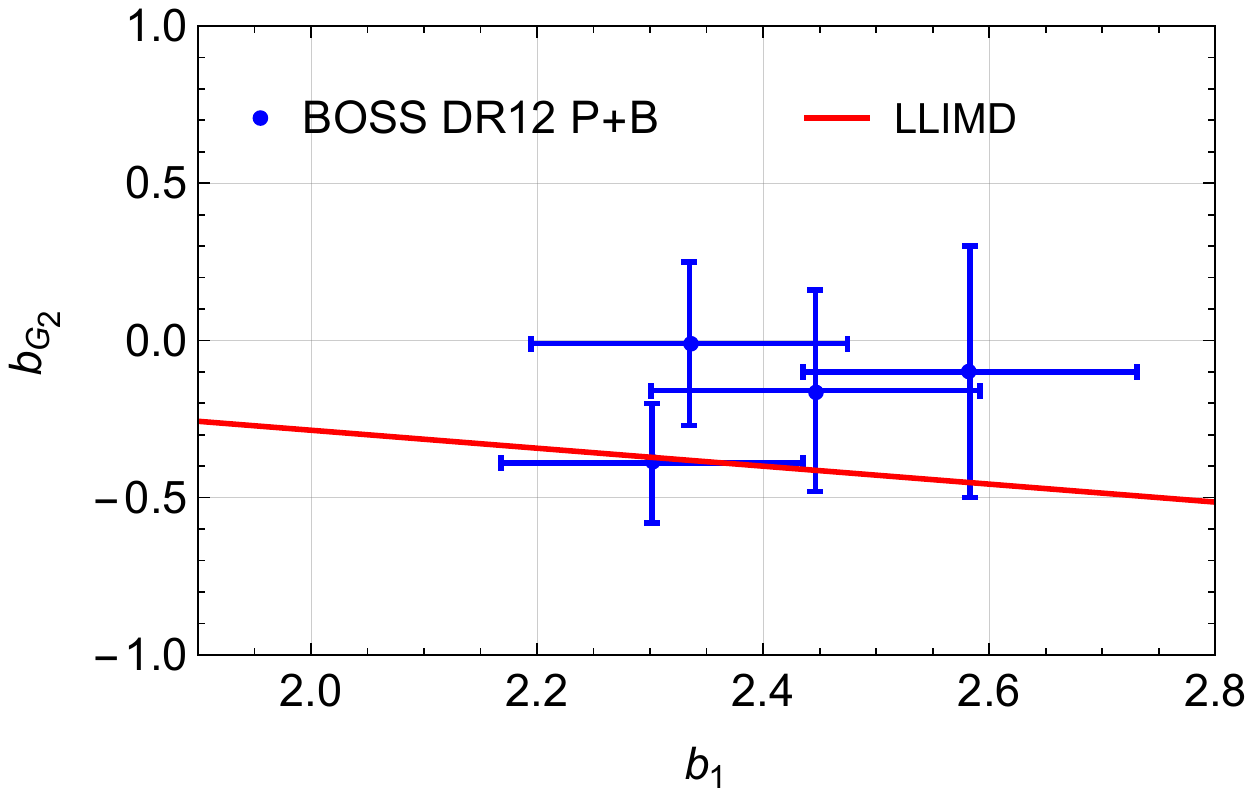}
	\caption{Linear, quadratic, and tidal bias parameters extracted from the analysis of BOSS DR12 power spectrum and bispectrum data, across two redshift slices and two sky regions. Data-points are shown in blue, whilst the red curves give predictions for the dark-matter bias relations, from \citep{2016JCAP...02..018L} and \citep{2018PhR...733....1D} for $b_2$ and $b_{\mathcal{G}_2}$ respectively. 
	Note that in our convention $b_2=b^{\text{ref. \citep{2016JCAP...02..018L}}}_2+\frac{4}{3}b_{\mathcal{G}_2}$.
	We find no evidence that galaxy bias parameters obey different relations to those of dark matter.}\label{fig:bias}
\end{figure}

In Fig.\,\ref{fig:bias} we plot the measured quadratic and tidal biases as a function of the linear bias, $b_1$, and compare this to a fit for $b_2(b_1)$ from dark matter halos \citep{2016JCAP...02..018L}, and the popular Lagrangian Local-in-Matter-Density (LLIMD) prediction for $b_{\mathcal{G}_2}(b_1)$. At the current level of precision, we do not detect any significant deviations, thus our results are consistent with the hypothesis that galaxy biases follow dark matter trends. Although some deviations have been found in HOD-enhanced $N$-body simulations \citep{2020PhRvD.102j3530E,Eggemeier:2021cam,2021arXiv211010161I} and hydrodynamic simulations \citep{2021JCAP...08..029B}, these are too small to detect in BOSS data, though this will likely change with the advent of DESI and Euclid. This also functions as an important consistency test: if the inclusion of the bispectrum had led to strong deviations from the dark matter relations, this would suggest that the bispectrum preferred very different bias parameters to those of the (well-tested) power spectrum. A generic prediction of perturbative models is that the power spectrum and bispectrum depend on the same set of biases, thus this would have indicated that the model was overfitting the data.





\section{Summary and Conclusions}\label{sec: summary}
In this work, we have presented a full analysis of the BOSS DR12 dataset, utilizing the power spectrum multipoles \citep{2020JCAP...05..042I}, the real-space power spectrum proxy \citep{2021arXiv211000006I}, BAO parameters obtained from the reconstructed power spectra \citep{2020JCAP...05..032P} and the bispectrum monopole. Unlike previous analyses, we have measured both the power spectrum and bispectrum using window-free estimators \citep{2021PhRvD.103j3504P,2021arXiv210706287P}, which allow them to be straightforwardly compared to theory, with no need for expensive convolution integrals. This enables the bispectrum to be analyzed in reasonable computational time without resorting to systematic-inducing approximations or large-scale cuts \citep[e.g.,][]{2017MNRAS.465.1757G}. The above work represents the first self-consistent analysis of the power spectrum and bispectrum, obtained using a robust theoretical model based on the Effective Field Theory of Large Scale Structure \citep{2021arXiv211010161I}.

By comparison with high-fidelity mock catalogs we have demonstrated the pipeline to be highly robust, and have used this to place the sharpest ever constraints on $\Lambda$CDM parameters from a full-shape analysis of BOSS DR12 and a BBN prior on the baryon density. In particular, we find $H_0$ constraints consistent with previous studies, somewhat below those of SH0ES, and obtain a $<5\%$ constraint on $\sigma_8$, equal to $0.722^{+0.032}_{-0.036}$ from the complete analysis with a \textit{Planck} prior on $n_s$. The inclusion of the bispectrum is found to significantly sharpen constraints on higher-order bias parameters, such as those controlling tidal physics, and leads to a $\approx 13\%$ reduction in the $\sigma_8$ error-bar, even with our highly conservative analysis choices. If one accepts a slightly larger systematic error budget, or reduces the physical priors on bias parameters, the bispectrum's utility will grow further. 

Overall, our results are consistent with those of \textit{Planck} at $95\%$ confidence level \citep{2020A&A...641A...6P}, but also broadly support the trend for lower $S_8$ measurements seen in weak lensing data-sets \citep[e.g.,][]{2021arXiv210513549D}, with $S_8 = 0.751\pm0.039$ found herein. However, we show that our constraint is driven by the large-scale quadrupole amplitude (with $k<0.1\hMpc$), which disfavors simplistic new physics explanations for the $S_8$ discrepancy. We have additionally compared our power spectrum results to standard windowed analyses, and those of other groups, and generally find good agreement. 

\vskip 4pt

It is interesting to compare the above conclusions to those of simplified power spectrum and bispectrum forecasts \citep{Yankelevich:2018uaz,2019JCAP...11..034C,Hahn:2019zob,2021JCAP...03..021A}. Perhaps the most relevant is that of \citep{2019JCAP...11..034C}, which performed a full MCMC forecast for the Euclid spectroscopic survey, utilizing a theoretical model based on perturbation theory (but without calibration to simulations). The former work suggested that inclusion of the tree-level bispectrum monopole would lead to significant improvements on a range of $\Lambda$CDM parameters (as well as the neutrino mass), whilst here we find appreciable changes only in $\sigma_8$. In part, this is caused by the difference in galaxy sample. \citep{2019JCAP...11..034C} focused on the Euclid or DESI-like emission line galaxies (ELGs), whilst we here consider only the BOSS luminous red galaxies (LRGs), which violate many of \citep{2019JCAP...11..034C}'s assumptions. For instance, preliminary analyses of the ELG power spectrum suggest that this sample has a weak fingers-of-God signature \cite{2021PhRvD.104j3514I}, closer to that assumed in \citep{2019JCAP...11..034C}. This directly impacts both the theoretical model and the assumed systematic error kernels, which can have a strong impact on the parameter inferences \cite[e.g.,][]{Chudaykin:2020hbf}. In contrast, LRGs exhibit strong fingers-of-God (velocity dispersion) effects, implying that the perturbative approximations break down at larger scales, limiting our $k$-range. Furthermore, the analysis of \citep{2019JCAP...11..034C} ignored a number of physical effects such as scale-dependent stochasticity; it is unclear whether these will be important for the DESI ELG sample, given that they appear to be present in the LRG sample \citep{2020PhRvD.102l3541N,2021JCAP...05..059S,2021arXiv211010161I}. All in all, our work does not invalidate the results of the previous forecast, but it remains to be seen whether their assumptions will be valid when analyzing the upcoming ELG samples. 


This work has demonstrated that robust power spectrum and bispectrum analyses are possible for current datasets, and will open the door to a number of future analyses, both extending the above scope, and testing new models of physics. Possibilities include:
\begin{itemize}
	\item A search for signatures of non-$\Lambda$CDM physics. In particular, we can obtain constraints on the neutrino mass using the above framework, sharpen bounds on Early Dark Energy \citep[e.g.][]{2019PhRvL.122v1301P,2020PhRvD.102j3502I}, and place constraints on primordial non-Gaussianity (PNG). The latter is a particular use-case for the bispectrum: whilst local-type PNG can be probed in the galaxy power spectrum \citep[e.g.,][]{2010CQGra..27l4011D}, features such as equilateral PNG can only be probed with higher-order statistics~\cite{Sefusatti:2009qh,Liguori:2010hx,MoradinezhadDizgah:2020whw}. Such extensions will be discussed in \resub{future work} \citep{nlpng-boss}.
	\item Extension to the anisotropic bispectrum multipoles, $B_{\ell}$ (or spherical harmonics, $B_{\ell m}$), analogous to the power spectrum multipoles $P_\ell$ \citep{2017MNRAS.467..928G,2017JCAP...12..020R,2020JCAP...06..041G}. Such quantities can be measured using a simple extension of the estimators discussed in \S\ref{sec: estimators}, and modeled simply by integrating the perturbation theory of \S\ref{sec: model} against a Legendre polynomial (or spherical harmonic). This will likely enhance the cosmological utility of the bispectrum at fixed $k_{\rm max}$, and reduce any systematics arising from anisotropy of the bispectrum window function.
	\item Implementation of a real-space bispectrum estimator, analogous to our $Q_0$ statistic \citep{2021arXiv211000006I} for the power spectrum. This would eliminate the necessity to model stochastic velocities, and increase the range of scales over which the bispectrum could be robustly modeled. An additional option would be to use a theoretical error model \citep[e.g.,][]{2016arXiv160200674B} to smooth over poorly scales, such that we can extract well-understood oscillatory information in the bispectrum up to high $k$ (cf.\,\citep{2020JCAP...05..032P} for reconstructed power spectra). This information is highly degenerate with that of the BAO parameters however.
	\item Development of a one-loop model for the galaxy bispectrum, and a more complete treatment of fingers-of-God \citep[e.g.,][]{2017JCAP...10..009H}, allowing a larger $k$-reach in both real- and redshift-space. As shown in \citep{2021arXiv211010161I}, redshift-space effects are a limiting factor in our current modeling, and forces us to use lower $k_{\rm max}$ than possible in real-space. Alternatively, we may consider map-level fingers-of-God compression \citep[e.g.,][]{2009ApJ...698..143R,2013JCAP...08..019H,2021JCAP...05..059S}, reducing the amplitude of the effect in the measured statistics.
	\item Compression of the dataset to reduce its dimensionality. This will be of use if narrower $k$-bins are used, or a larger $k_{\rm max}$, and will help reduce the noise penalty from using a finite number of mocks to derive Gaussian covariance matrices (cf.\,\ref{subsec: pk-analysis-comparison}). This may proceed via a generic linear approach \citep[e.g.,][]{2021PhRvD.103d3508P}, or a bispectrum specific framework \citep{2019MNRAS.484.3713G,2019MNRAS.484L..29G,2019MNRAS.484..364S,2018MNRAS.476.4045G}, and the resulting coefficients can likely be estimated directly from the data \citep{2021arXiv210706287P}.
\end{itemize}

In addition, the current methodology can be reapplied to upcoming datasets such as that from DESI and Euclid. Given that the estimators are publicly available, this can be performed with relative ease, especially given that the theoretical model and window-free implementations have already been tested on survey volumes larger than that of the full DESI sample (cf.\,\citep{2021arXiv211010161I} and \S\ref{sec: consistency-tests}). Upcoming surveys will include multiple different galaxy samples: of particular interest are the emission line galaxy samples, since they are expected to have a lower contribution from stochastic velocities \citep{2021PhRvD.104j3514I}. This reduces the fingers-of-God effect, enabling smaller scales to be modeled, and thus obtain a greater volume of information from the bispectrum. 

As the data volume increases, it will become increasingly important to understand systematics of the data and analysis pipeline. In particular, an important part of any DESI analysis will be a rigorous study of observational effects, such as those arising from sky calibration and poorly known redshift distributions. Given that our analyses include information from a broader range of scales than traditional methods, we are more susceptible to such errors. Assuming that these phenomena can be controlled, we may proceed to perform analysis of upcoming data with higher-order statistics, and thus reap the corresponding rewards.

\begin{acknowledgments}
\footnotesize
We are indebted to Giovanni Cabass, Marko Simonovic, and Matias Zaldarriaga for a wealth of insightful discussions. We are additionally grateful to Kazuyuki Akitsu, \resub{Shi-Fan Chen}, Simone Ferraro, Yosuke Kobayashi, Pat McDonald, Takahiro Nishimichi, David Spergel, Masahiro Takada and Martin White for useful feedback, \resub{and the anonymous referee for a insightful comments}. OHEP thanks the Simons Foundation for additional support. The work of MMI has been supported by NASA through the NASA Hubble Fellowship grant \#HST-HF2-51483.001-A awarded by the Space Telescope Science Institute, which is operated by the Association of Universities for Research in Astronomy, Incorporated, under NASA contract NAS5-26555.

\vskip 2pt

The authors acknowledge that the work reported in this paper was substantially performed using the Research Computing resources at Princeton University which is a consortium of groups led by the Princeton Institute for Computational Science and Engineering (PICSciE) and the Office of Information Technology's Research Computing Division.

Funding for SDSS-III has been provided by the Alfred P. Sloan Foundation, the Participating Institutions, the National Science Foundation, and the U.S. Department of Energy Office of Science. The SDSS-III web site is \href{http://www.sdss3.org/}{www.sdss3.org}.

SDSS-III is managed by the Astrophysical Research Consortium for the Participating Institutions of the SDSS-III Collaboration including the University of Arizona, the Brazilian Participation Group, Brookhaven National Laboratory, Carnegie Mellon University, University of Florida, the French Participation Group, the German Participation Group, Harvard University, the Instituto de Astrofisica de Canarias, the Michigan State/Notre Dame/JINA Participation Group, Johns Hopkins University, Lawrence Berkeley National Laboratory, Max Planck Institute for Astrophysics, Max Planck Institute for Extraterrestrial Physics, New Mexico State University, New York University, Ohio State University, Pennsylvania State University, University of Portsmouth, Princeton University, the Spanish Participation Group, University of Tokyo, University of Utah, Vanderbilt University, University of Virginia, University of Washington, and Yale University. 

The massive production of all MultiDark-Patchy mocks for the BOSS Final Data Release has been performed at the BSC Marenostrum supercomputer, the Hydra cluster at the Instituto de Fısica Teorica UAM/CSIC, and NERSC at the Lawrence Berkeley National Laboratory. We acknowledge support from the Spanish MICINNs Consolider-Ingenio 2010 Programme under grant MultiDark CSD2009-00064, MINECO Centro de Excelencia Severo Ochoa Programme under grant SEV- 2012-0249, and grant AYA2014-60641-C2-1-P. The MultiDark-Patchy mocks was an effort led from the IFT UAM-CSIC by F. Prada’s group (C.-H. Chuang, S. Rodriguez-Torres and C. Scoccola) in collaboration with C. Zhao (Tsinghua U.), F.-S. Kitaura (AIP), A. Klypin (NMSU), G. Yepes (UAM), and the BOSS galaxy clustering working group.

\end{acknowledgments}

\appendix 

\section{Full Parameter Constraints}\label{appen: full-constraints}
\begin{figure}
    \centering
    \includegraphics[width=\textwidth]{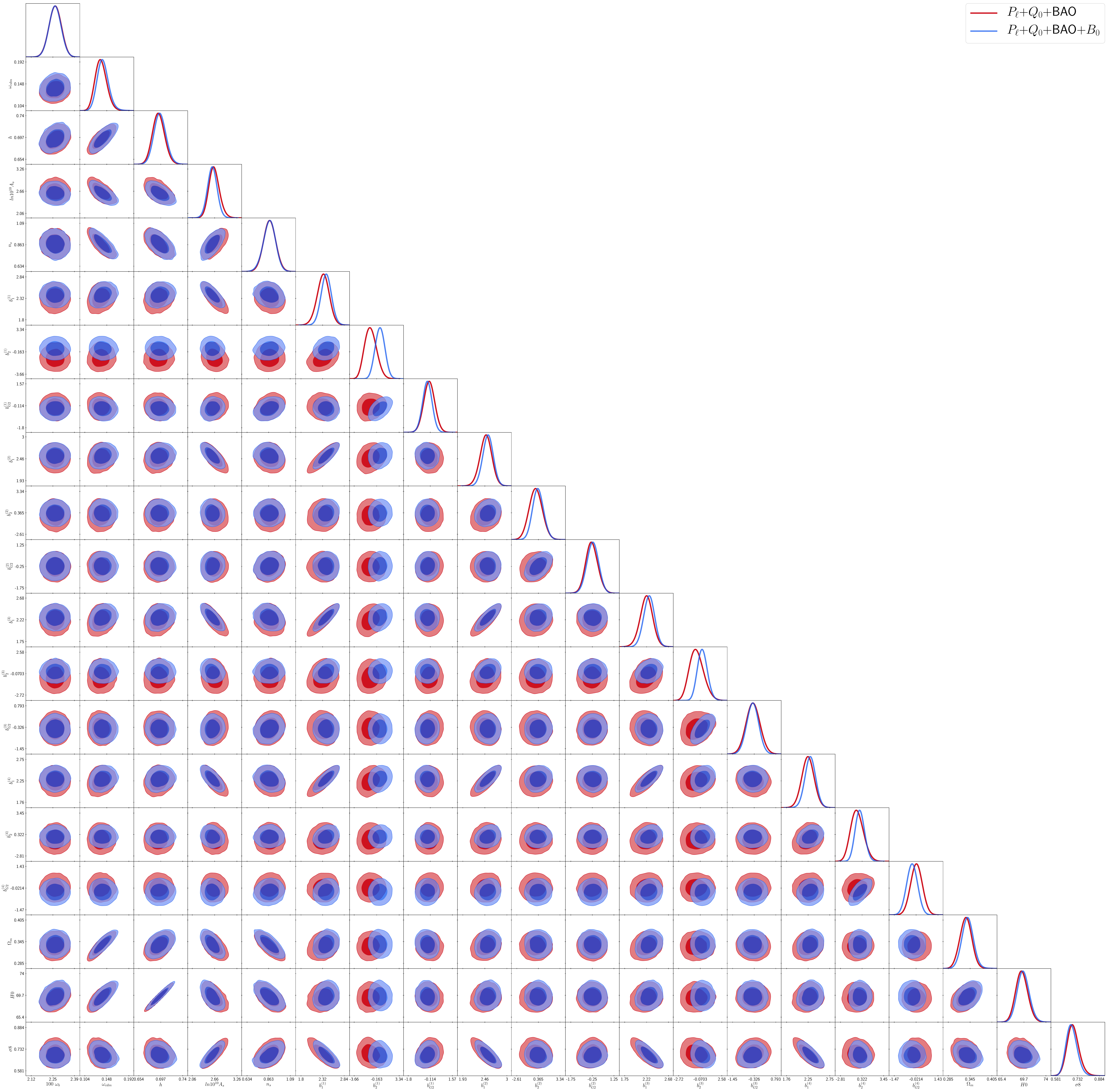}
    \caption{Full posterior plot of the cosmological and nuisance parameter posteriors measured from the BOSS DR12 data, with a BBN prior on the $\omega_b$. We show results for both the full power spectrum ($P_\ell + Q_0 + \mathrm{BAO}$) and the joint analysis of power spectrum and bispectrum. The corresponding parameter constraints are given in Tab.\,\ref{tab: all-constraints-free-ns}.}
    \label{fig: all-corner-free-ns}
\end{figure}

\begin{figure}
    \centering
    \includegraphics[width=\textwidth]{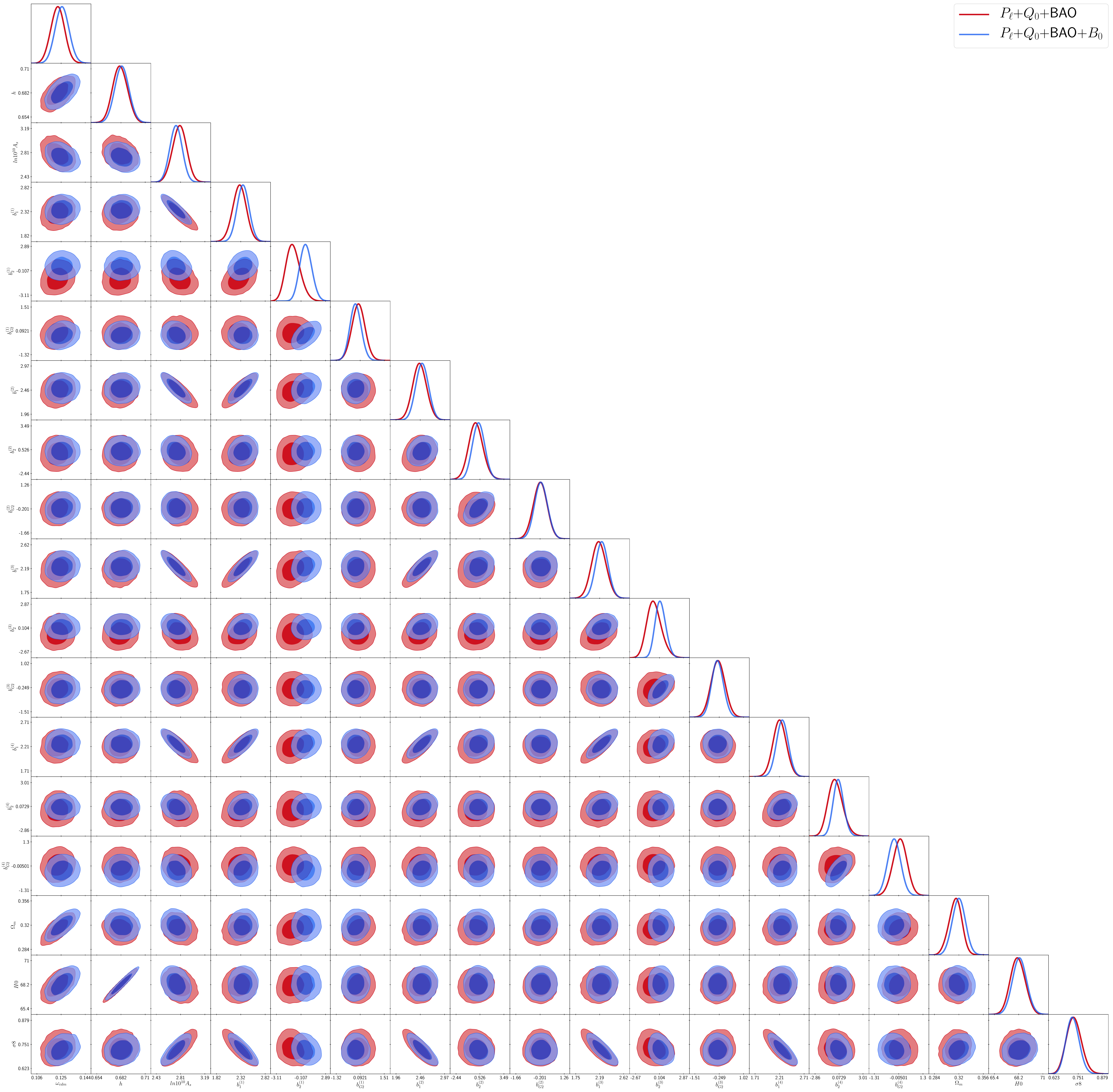}
    \caption{As Fig.\,\ref{fig: all-corner-free-ns}, but including a \textit{Planck} prior on $n_s$. The corresponding parameter constraints are given in Tab.\,\ref{tab: all-constraints-fix-ns}.}
    \label{fig: all-corner-fix-ns}
\end{figure}

In this appendix, we give the full marginalized posteriors for all cosmological and non-cosmological parameters sampled within the MCMC likelihoods, using the full power spectrum likelihood (including BAO and $Q_0$ datasets), and the power spectrum plus bispectrum joint likelihood. Parameters entering the model linearly (such as $P_{\rm shot}$) are marginalized over analytically, and thus excluded. Results are shown in Tabs.\,\ref{tab: all-constraints-free-ns}\,\&\,\ref{tab: all-constraints-fix-ns}, and in Figs.\,\ref{fig: all-corner-free-ns}\,\&\,\ref{fig: all-corner-fix-ns}, with the latter set including a \textit{Planck} prior on $n_s$.

\begin{table}
    \centering
   \rowcolors{2}{white}{vlightgray}
  \begin{tabular}{|c|cccc|cccc|}
 \hline
 & \multicolumn{4}{c|}{$\bm{P_\ell + Q_0 + \mathrm{BAO}}$} & \multicolumn{4}{c|}{$\bm{P_\ell + Q_0 + \mathrm{BAO} + B_0}$}\\\hline
\quad \textbf{Parameter}\quad\quad & best-fit & mean$\,\pm\,\sigma$ & \quad 95\% lower \quad & \quad 95\% upper \quad & best-fit & mean$\,\pm\,\sigma$ & \quad 95\% lower \quad & \quad 95\% upper \quad \\ \hline
$\omega_{cdm }$ &$0.1322$ & $0.1366_{-0.013}^{+0.011}$ & $0.1139$ & $0.1601$ &$0.1378$ & $0.1405_{-0.013}^{+0.011}$ & $0.1175$ & $0.164$ \\
$h$ &$0.6903$ & $0.6931_{-0.012}^{+0.011}$ & $0.6688$ & $0.717$ &$0.6944$ & $0.6957_{-0.013}^{+0.011}$ & $0.6719$ & $0.7203$ \\
$\mathrm{ln}\left(10^{10}A_{s}\right)$ &$2.729$ & $2.657_{-0.15}^{+0.14}$ & $2.366$ & $2.951$ &$2.603$ & $2.599_{-0.14}^{+0.13}$ & $2.339$ & $2.871$ \\
$n_{s }$ &$0.891$ & $0.8741_{-0.064}^{+0.067}$ & $0.7431$ & $1.006$ &$0.8698$ & $0.8697_{-0.064}^{+0.067}$ & $0.7391$ & $1.003$ \\
$b^{(1)}_{1 }$ &$2.263$ & $2.331_{-0.15}^{+0.15}$ & $2.031$ & $2.644$ &$2.406$ & $2.411_{-0.13}^{+0.13}$ & $2.154$ & $2.67$ \\
$b^{(1)}_{2 }$ &$-0.9937$ & $-1.094_{-1}^{+0.84}$ & $-2.881$ & $0.8157$ &$0.1082$ & $0.3584_{-0.78}^{+0.71}$ & $-1.124$ & $1.881$ \\
$b^{(1)}_{{\mathcal{G}_2} }$ &$-0.2681$ & $-0.1914_{-0.43}^{+0.43}$ & $-1.047$ & $0.6657$ &$-0.4004$ & $-0.3368_{-0.37}^{+0.37}$ & $-1.089$ & $0.3987$ \\
$b^{(2)}_{1 }$ &$2.411$ & $2.483_{-0.15}^{+0.15}$ & $2.186$ & $2.786$ &$2.564$ & $2.539_{-0.13}^{+0.13}$ & $2.278$ & $2.797$ \\
$b^{(2)}_{2 }$ &$-0.3943$ & $-0.004331_{-0.93}^{+0.91}$ & $-1.812$ & $1.853$ &$0.7517$ & $0.3415_{-0.8}^{+0.75}$ & $-1.183$ & $1.901$ \\
$b^{(2)}_{{\mathcal{G}_2} }$ &$-0.6292$ & $-0.2781_{-0.44}^{+0.43}$ & $-1.144$ & $0.5858$ &$-0.3779$ & \quad $-0.2246_{-0.41}^{+0.41}$ & $-1.045$ & $0.5843$ \\
$b^{(3)}_{1 }$ &$2.15$ & $2.211_{-0.14}^{+0.14}$ & $1.932$ & $2.489$ &$2.239$ & $2.269_{-0.12}^{+0.12}$ & $2.031$ & $2.504$ \\
$b^{(3)}_{2 }$ &$-0.3785$ & $-0.509_{-0.97}^{+0.79}$ & $-2.189$ & $1.301$ &$0.2004$ & $0.2082_{-0.64}^{+0.59}$ & $-1.018$ & $1.438$ \\
$b^{(3)}_{{\mathcal{G}_2} }$ &$-0.3157$ & $-0.3832_{-0.37}^{+0.37}$ & $-1.114$ & $0.3501$ &$-0.3576$ & $-0.4074_{-0.32}^{+0.32}$ & $-1.044$ & $0.2281$ \\
$b^{(4)}_{1 }$ &$2.176$ & $2.248_{-0.14}^{+0.14}$ & $1.961$ & $2.532$ &$2.307$ & $2.303_{-0.12}^{+0.12}$ & $2.056$ & $2.552$ \\
$b^{(4)}_{2 }$ &$-0.2319$ & $-0.3743_{-1}^{+0.87}$ & $-2.178$ & $1.523$ &$0.042$ & $0.006909_{-0.72}^{+0.65}$ & $-1.346$ & $1.381$ \\
$b^{(4)}_{{\mathcal{G}_2} }$ &$-0.05649$ & \quad$-0.002389_{-0.4}^{+0.38}$ & $-0.7803$ & $0.7901$ &$-0.219$ & $-0.2876_{-0.37}^{+0.37}$ & $-1.018$ & $0.4428$ \\\hline
$\Omega_{m }$ &$0.327$ & $0.3326_{-0.018}^{+0.016}$ & $0.2998$ & $0.366$ &$0.3342$ & $0.3381_{-0.017}^{+0.016}$ & $0.3056$ & $0.3703$ \\
$H_0$ &$69.03$ & $69.31_{-1.2}^{+1.1}$ & $66.88$ & $71.7$ &$69.44$ & $69.57_{-1.3}^{+1.1}$ & $67.19$ & $72.03$ \\
$\sigma_8$ &$0.7158$ & $0.7011_{-0.045}^{+0.04}$ & $0.6169$ & $0.7876$ &$0.6856$ & $0.6917_{-0.041}^{+0.035}$ & $0.6165$ & $0.7698$ \\
\hline
 \end{tabular}
 \caption{Full parameter constraints from the $\Lambda$CDM analysis of BOSS DR12 data using the power spectrum datasets ($P_\ell$+$Q_0$+BAO, left) and including the bispectrum ($P_\ell$+$Q_0$+BAO, right). We give the best-fit values, the mean, 68\%, and 95\% confidence level results in each case, and show the derived parameters in the bottom rows. The superscripts on bias parameters indicate the sample, in the order NGC z3, SGC z3, NGC z1, SGC z1. The associated two-dimensional posteriors are shown in Fig.\,\ref{fig: all-corner-free-ns}. Corresponding results with a \textit{Planck} prior on $n_s$ are shown in Tab.\,\ref{tab: all-constraints-fix-ns}.}\label{tab: all-constraints-free-ns}
\end{table}

\begin{table}
    \centering
   \rowcolors{2}{white}{vlightgray}
  \begin{tabular}{|c|cccc|cccc|}
  \hline
  & \multicolumn{4}{c|}{$\bm{P_\ell + Q_0 + \mathrm{BAO}}$} & \multicolumn{4}{c|}{$\bm{P_\ell + Q_0 + \mathrm{BAO} + B_0}$}\\\hline
  \quad \textbf{Parameter}\quad\quad & best-fit & mean$\,\pm\,\sigma$ & \quad 95\% lower \quad & \quad 95\% upper \quad & best-fit & mean$\,\pm\,\sigma$ & \quad 95\% lower \quad & \quad 95\% upper \quad \\ \hline
$\omega_{\rm cdm }$ &$0.1218$ & $0.1227_{-0.0059}^{+0.0053}$ & $0.1118$ & $0.134$ &$0.1242$ & $0.1262_{-0.0059}^{+0.0053}$ & $0.1152$ & $0.1374$ \\
$h$ &$0.6778$ & $0.6811_{-0.0089}^{+0.0083}$ & $0.6641$ & $0.6981$ &$0.6809$ & $0.6831_{-0.0086}^{+0.0083}$ & $0.6665$ & $0.7002$ \\
$ln\left(10^{10}A_{s }\right)$ &$2.863$ & $2.795_{-0.12}^{+0.11}$ & $2.572$ & $3.022$ &$2.771$ & $2.741_{-0.098}^{+0.096}$ & $2.548$ & $2.935$ \\
$b^{(1)}_{1 }$ &$2.217$ & $2.288_{-0.15}^{+0.15}$ & $1.993$ & $2.579$ &$2.335$ & $2.365_{-0.13}^{+0.12}$ & $2.115$ & $2.619$ \\
$b^{(1)}_{2 }$ &$-1.033$ & $-1.045_{-0.96}^{+0.81}$ & $-2.769$ & $0.74$ &$0.4944$ & $0.4867_{-0.78}^{+0.69}$ & $-0.9661$ & $1.976$ \\
$b^{(1)}_{{\mathcal{G}_2} }$ &\quad$-0.03938$ & \quad$-0.01572_{-0.41}^{+0.39}$ & $-0.8159$ & $0.7855$ &$-0.08666$ & $-0.1783_{-0.34}^{+0.34}$ & $-0.8567$ & $0.5097$ \\
$b^{(2)}_{1 }$ &$2.387$ & $2.449_{-0.14}^{+0.15}$ & $2.156$ & $2.744$ &$2.471$ & $2.502_{-0.13}^{+0.13}$ & $2.243$ & $2.766$ \\
$b^{(2)}_{2 }$ &$-0.4042$ & $0.01267_{-0.98}^{+0.88}$ & $-1.793$ & $1.868$ &$0.5985$ & $0.3589_{-0.79}^{+0.74}$ & $-1.16$ & $1.886$ \\
$b^{(2)}_{{\mathcal{G}_2} }$ &$-0.3214$ & $-0.1954_{-0.42}^{+0.42}$ & $-1.03$ & $0.6524$ &$-0.2001$ & $-0.1392_{-0.4}^{+0.39}$ & $-0.9232$ & $0.6532$ \\
$b^{(3)}_{1 }$ &$2.093$ & $2.172_{-0.13}^{+0.13}$ & $1.9$ & $2.44$ &$2.179$ & $2.227_{-0.12}^{+0.11}$ & $1.996$ & $2.459$ \\
$b^{(3)}_{2 }$ &$-0.5351$ & $-0.4924_{-0.96}^{+0.76}$ & $-2.168$ & $1.276$ &$0.2993$ & $0.2286_{-0.64}^{+0.58}$ & $-0.9688$ & $1.469$ \\
$b^{(3)}_{{\mathcal{G}_2} }$ &$-0.4988$ & $-0.3075_{-0.37}^{+0.36}$ & $-1.026$ & $0.4293$ &$-0.335$ & $-0.3311_{-0.32}^{+0.3}$ & $-0.9516$ & $0.2952$ \\
$b^{(4)}_{1 }$ &$2.141$ & $2.209_{-0.14}^{+0.14}$ & $1.933$ & $2.489$ &$2.207$ & $2.264_{-0.13}^{+0.12}$ & $2.022$ & $2.509$ \\
$b^{(4)}_{2 }$ &$-0.8938$ & $-0.358_{-1}^{+0.81}$ & $-2.106$ & $1.486$ &$-0.4185$ & \quad$0.007424_{-0.71}^{+0.63}$ & $-1.293$ & $1.349$ \\
$b^{(4)}_{{\mathcal{G}_2} }$ &$0.02433$ & $0.05553_{-0.39}^{+0.37}$ & $-0.7031$ & $0.8315$ &$-0.3927$ & $-0.2495_{-0.36}^{+0.35}$ & $-0.9517$ & $0.4648$ \\\hline
$\Omega_{m }$ &$0.3153$ & $0.3142_{-0.01}^{+0.0095}$ & $0.2949$ & $0.3338$ &$0.3176$ & $0.3197_{-0.01}^{+0.0095}$ & $0.3005$ & $0.3393$ \\
$H_0$ &$67.78$ & $68.11_{-0.89}^{+0.83}$ & $66.41$ & $69.81$ &$68.09$ & $68.31_{-0.86}^{+0.83}$ & $66.65$ & $70.02$ \\
$\sigma_8$ &$0.7491$ & $0.7286_{-0.042}^{+0.036}$ & $0.6511$ & $0.8088$ &$0.7248$ & $0.722_{-0.036}^{+0.032}$ & $0.6536$ & $0.7915$ \\\hline
 \end{tabular}
 \caption{As Tab.\,\ref{tab: all-constraints-free-ns}, but including a \textit{Planck} prior on the spectral slope $n_s$. The associated two-dimensional posteriors are shown in Fig.\,\ref{fig: all-corner-fix-ns}.}\label{tab: all-constraints-fix-ns}
\end{table}

\bibliographystyle{JHEP}
\bibliography{oliverlib,mishalib}%

\end{document}